\documentclass[12pt]{article}
\pdfoutput=1

\usepackage{scicite}

\usepackage{times}

\usepackage{graphicx}

\usepackage{multicol}

\newenvironment{sciabstract}{%
\begin{quote} \bf}
{\end{quote}}

\newcounter{lastnote}

\usepackage{authblk}
\usepackage{xcolor}
\usepackage[colorlinks=true,linkcolor=blue,urlcolor=blue,citecolor=blue,bookmarksopenlevel=2,bookmarksopen=true]{hyperref}
\usepackage[labelfont=bf]{caption}
\usepackage{amsmath}
\usepackage{amssymb}
\usepackage{aas_macros_long}
\usepackage{booktabs}
\usepackage{lscape}
\usepackage{longtable}
\usepackage{threeparttablex}
\usepackage{xspace}
\newcommand{\msun}{{\rm M}_{\odot}}
\newcommand{\lsun}{{\rm L}_{\odot}}
\newcommand{\rsun}{{\rm R}_{\odot}}

\newcommand{\kms}{{\rm km\,s^{-1}}}
\newcommand{\myr}{{\rm Myr}}
\newcommand{\bonnsai}{\mbox{\textsc{Bonnsai}}\xspace}
\newcommand{\fastwind}{\mbox{\textsc{Fastwind}}\xspace}

\newcommand{\tdor}{\mbox{30~Dor}\xspace}
\newcommand{\flames}{\mbox{FLAMES}\xspace}
\newcommand{\argus}{\mbox{ARGUS}\xspace}

\newcommand{\pulsar}{\mbox{PSR~J0537-6910}\xspace}
\newcommand{\eg}{e.g.\@\xspace}
\newcommand{\ie}{i.e.\@\xspace}
\newcommand{\V}{\emph{V}\@\xspace}

\newcommand{\mytitle}{An excess of massive stars in the local 30~Doradus starburst}
\title{\mytitle\footnote{This is the authors' version. The definitive version is published in Science on 5th Jan 2018: Vol.\ 359, Issue 6371, pp.\ 69-71 DOI: 10.1126/science.aan0106.}} % 96 chars including spaces

\author[1$\ast$]{F.R.N. Schneider}
\author[2]{H. Sana}
\author[3]{C.J. Evans}
\author[4,5]{J.M. Bestenlehner}
\author[6]{N. Castro}
\author[7]{L. Fossati}
\author[8]{G. Gr{\"a}fener}
\author[8]{N. Langer}
\author[3]{O.H. Ram{\'i}rez-Agudelo}
\author[9]{C. Sab{\'i}n-Sanjuli{\'a}n}
\author[10,11]{S. Sim{\'o}n-D{\'i}az}
\author[12]{F. Tramper}
\author[5]{P.A. Crowther}
\author[13,2]{A. de Koter}
\author[13]{S.E. de Mink}
\author[14]{P.L. Dufton}
\author[15]{M. Garcia}
\author[16]{M. Gieles}
\author[17,18]{V. H\'{e}nault-Brunet}
\author[10,11]{A. Herrero}
\author[19,16]{R.G. Izzard}
\author[20]{V. Kalari}
\author[12]{D.J. Lennon}
\author[21]{J. Ma\'{i}z Apell\'{a}niz}
\author[22]{N. Markova}
\author[15]{F. Najarro}
\author[1,8]{Ph. Podsiadlowski}
\author[23]{J. Puls}
\author[3]{W.D. Taylor}
\author[24]{J.Th. van Loon}
\author[25]{J.S. Vink}
\author[26,27]{C. Norman}

\affil[1]{\normalsize{Department of Physics, University of Oxford, Keble Rd, Oxford OX1 3RH, United Kingdom}}
\affil[2]{\normalsize{Institute of Astrophysics, KU Leuven, Celestijnenlaan 200D, 3001, Leuven, Belgium}}
\affil[3]{\normalsize{UK Astronomy Technology Centre, Royal Observatory Edinburgh, Blackford Hill, Edinburgh EH9 3HJ, United Kingdom}}
\affil[4]{\normalsize{Max-Planck-Institut f{\"u}r Astronomie, K{\"o}nigstuhl 17, 69117 Heidelberg, Germany}}
\affil[5]{\normalsize{Department of Physics and Astronomy, Hicks Building, Hounsfield Road, University of Sheffield, Sheffield S3 7RH, United Kingdom}}
\affil[6]{\normalsize{Department of Astronomy, University of Michigan, 1085 S. University Avenue, Ann Arbor, MI 48109-1107, USA}}
\affil[7]{\normalsize{Austrian Academy of Sciences, Space Research Institute, Schmiedlstra{\ss}e 6, 8042 Graz, Austria}}
\affil[8]{\normalsize{Argelander-Institut f{\"u}r Astronomie der Universit{\"a}t Bonn, Auf dem H{\"u}gel~71, 53121~Bonn, Germany}}
\affil[9]{\normalsize{Departamento de F{\'i}sica y Astronom{\'i}a, Universidad de La Serena, Avda. Juan Cisternas $N^o$ 1200 Norte, La Serena, Chile}}
\affil[10]{\normalsize{Instituto de Astrof{\'i}sica de Canarias, E-38205 La Laguna, Tenerife, Spain}}
\affil[11]{\normalsize{Departamento de Astrof{\'i}sica, Universidad de La Laguna, E-38206 La Laguna, Tenerife, Spain}}
\affil[12]{\normalsize{European Space Astronomy Centre, Mission Operations Division, PO Box 78, 28691 Villanueva de la Ca\~nada, Madrid, Spain}}
\affil[13]{\normalsize{Astronomical Institute Anton Pannekoek, Amsterdam University, Science Park 904, 1098 XH Amsterdam, The Netherlands}}
\affil[14]{\normalsize{Astrophysics Research Centre, School of Mathematics and Physics, Queen's University Belfast, Belfast BT7 1NN, Northern Ireland, United Kingdom}}
\affil[15]{\normalsize{Centro de Astrobiolog\'ia (CSIC-INTA), Ctra. de Torrej\'on a Ajalvir km-4, E-28850 Torrej\'on de Ardoz, Madrid, Spain}}
\affil[16]{\normalsize{Department of Physics, Faculty of Engineering and Physical Sciences, University of Surrey, Guildford, GU2 7XH, United Kingdom}}
\affil[17]{\normalsize{National Research Council, Herzberg Astronomy \& Astrophysics, 5071 West Saanich Road, Victoria, BC, V9E 2E7, Canada}}
\affil[18]{\normalsize{Department of Astrophysics/IMAPP, Radboud University, PO Box 9010, NL-6500 GL Nijmegen, The Netherlands}}
\affil[19]{\normalsize{Institute of Astronomy, The Observatories, Madingley Road, Cambridge CB3 0HA, United Kingdom}}
\affil[20]{\normalsize{Departamento de Astronom{\'i}a, Universidad de Chile, Camino El Observatorio 1515, Las Condes, Santiago, Casilla 36-D, Chile}}
\affil[21]{\normalsize{Centro de Astrobiolog{\'i}a, CSIC-INTA, ESAC campus, camino bajo del castillo s/n, E-28\,692 Villanueva de la Ca\~nada, Spain}}
\affil[22]{\normalsize{Institute of Astronomy with National Astronomical Observatory, Bulgarian Academy of Sciences, PO Box 136, 4700 Smoljan, Bulgaria}}
\affil[23]{\normalsize{Ludwig-Maximilians-Universit{\"a}t M{\"u}nchen, Universit{\"a}tssternwarte, Scheinerstrasse 1, 81679 M{\"u}nchen, Germany}}
\affil[24]{\normalsize{Lennard-Jones Laboratories, Keele University, Staffordshire, ST5 5BG, United Kingdom}}
\affil[25]{\normalsize{Armagh Observatory, College Hill, Armagh, BT61 9DG, Northern Ireland, United Kingdom}}
\affil[26]{\normalsize{Johns Hopkins University, Homewood Campus, Baltimore, MD 21218, USA}}
\affil[27]{\normalsize{Space Telescope Science Institute, 3700 San Martin Drive, Baltimore, MD 21218, USA}\vspace{0.5cm}}

\affil[$\ast$]{\normalsize{To whom correspondence should be addressed; E-mail: fabian.schneider@physics.ox.ac.uk.}}

\date{}

\begin{document}

%\baselineskip24pt

\maketitle 


% 125 words or less, no citations and abbreviations; 
\begin{sciabstract}
The 30\,Doradus star-forming region in the Large Magellanic Cloud is a 
nearby analogue of large star-formation events in the distant Universe.
We determine the recent formation history and 
the initial mass function (IMF) of massive stars in 30\,Doradus 
based on spectroscopic observations of 247 stars more massive than 15 solar masses ($\msun$). 
The main episode of massive star formation started 
about $8\,\mathrm{Myr}$ ago and the star-formation rate seems to have declined in the last $1\,\mathrm{Myr}$. 
The IMF is densely sampled up to $200\,\msun$ and 
contains $32\pm12\%$ more stars above $30\,\msun$ than predicted by a standard Salpeter IMF.
In the mass range $15\text{--}200\,\msun$, the IMF power-law exponent is $1.90^{+0.37}_{-0.26}$, 
shallower than the Salpeter value of 2.35.
\end{sciabstract}

\begin{multicols}{2}
% Main body of paper
Starbursts are large star-formation events whose feedback affects
the dynamical and chemical evolution of star-forming galaxies
throughout cosmic history \cite{2012ARA&A..50..531K,2014ARA&A..52..415M,2009ApJ...695..292C}.  
They are found at low and high redshift, with the earliest starburst galaxies contributing to
the reionisation of the Universe \cite{2001ARA&A..39...19L,2014ARA&A..52..415M}. 
In such starbursts, massive stars ($\geq 10\,\msun$) dominate 
the feedback through intense ionising radiation, stellar outflows and supernova explosions. 
Because of large distances to most starbursts, analyses have so far been restricted either to
photometric observations or to composite spectra of
entire stellar populations. In the former case, the high surface 
temperature of massive stars precludes the determination of accurate physical 
parameters because their colours are too similar \cite{1988ApJ...328..704H} and, in the latter case, physical 
parameters of individual stars cannot be determined \cite{1998A&A...329..409M}.
Greater understanding can be obtained by spectroscopically examining individual stars within star-forming regions.

The IMF influences many areas of astrophysics because it determines the 
relative fraction of massive stars, \ie, those which undergo
supernova explosions and drive the evolution of star-forming galaxies.
Much effort has therefore gone into understanding whether the IMF is universal
or varies with local environmental properties \cite{2010ARA&A..48..339B,2017MNRAS.464.1738D}. 
Over the last few decades, evidence has accumulated that 
the IMF slope may be flatter than that of a Salpeter IMF \cite{1955ApJ...121..161S}, 
\ie there are more high-mass stars than expected, in regions of intense star formation \cite{2005MNRAS.356.1191B,2011MNRAS.415.1647G,2012MNRAS.422.2246M}.
However, these studies are based on integrated properties of stellar populations,
hampering the ability to infer IMFs.

The star-forming region 30\,Doradus (\tdor) lies within the Large Magellanic Cloud (LMC), 
a satellite galaxy of the Milky Way, and has a metallicity (total abundance of all elements heavier than helium) 
of about 40\% the solar value \cite{2015ApJ...806...21D}.
At a distance of 50 kiloparsecs \cite{2013Natur.495...76P},
\tdor is a nearby analogue of distant starbursts 
and one of 
the brightest hydrogen-ionisation (H~\textsc{ii}) regions in the local Universe \cite{1988AJ.....95..720K}. 
With a diameter of about 200 parsecs, \tdor hosts several star clusters and associations, 
and is similar in size to luminous H~\textsc{ii} complexes in more distant galaxies \cite{2003AJ....126.2317O}. 

Using the Fibre Large Array Multi Element Spectrograph (\flames) \cite{2002Msngr.110....1P} 
on the Very Large Telescope (VLT), the VLT-FLAMES Tarantula Survey (VFTS) \cite{2011A&A...530A.108E} 
has obtained optical spectra of about 800 massive stars in \tdor, avoiding the core region
of the dense star cluster R136 because of difficulties with crowding \cite{2011A&A...530A.108E}. 
Repeated observations at multiple epochs allow determination of the orbital motion of 
potentially binary objects. For a sample of 452 apparently single stars, robust stellar parameters
such as effective temperatures, luminosities, surface gravities and projected rotational velocities
are found by modelling the observed spectra \cite{Methods}.
Composite spectra of visual multiple systems and spectroscopic binaries are not considered here because their 
parameters cannot be reliably inferred from the VFTS data.

% Inferred SFH and IMF
\begin{figure*}
\centering
\includegraphics[width=0.58\textwidth]{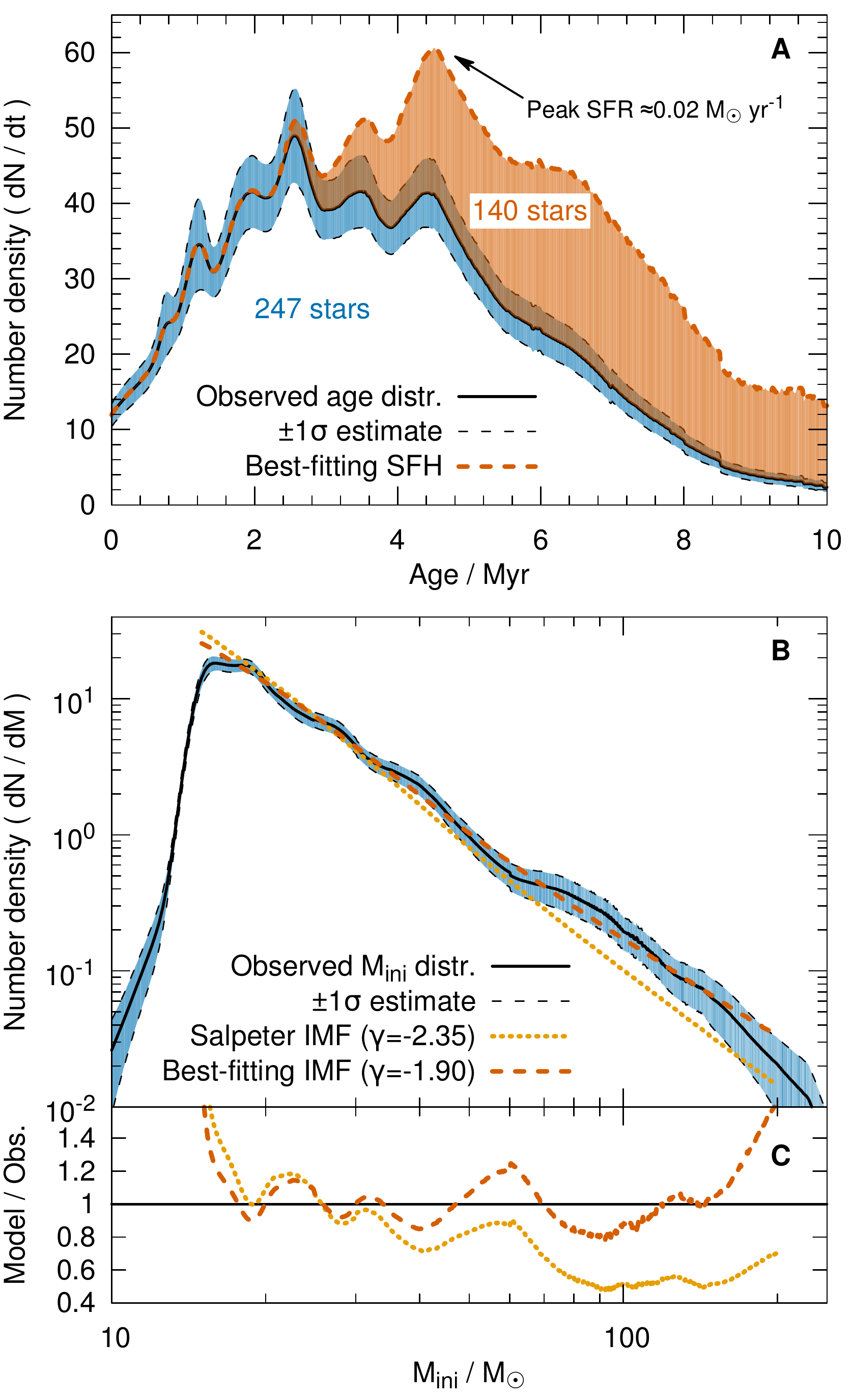}
\caption{\textbf{Age (A) and initial-mass, $M_\mathrm{ini}$, (B) distribution of the VFTS sample stars more massive than $15\,\msun$ (black line).}
Uncertainties are calculated by bootstrapping \cite{Methods} and the $1\sigma$ region is shaded blue. The best-fitting star-formation history (A) and present-day distribution of initial masses (B) are plotted in red. For comparison, also the expected present-day distribution of initial masses assuming a Salpeter IMF is provided (B; note that these modelled mass distributions are not single power-law functions anymore). About 140 stars above $15\,\msun$ are inferred to have ended their nuclear burning during the last $\approx 10\,\myr$ and their contribution to the SFH is shown by the red shaded region in panel (A). The peak star-formation rate (SFR) extrapolated to the whole \tdor region is about $0.02\,\msun\,\mathrm{yr}^{-1}$ (of order $\approx1\,\msun\,\mathrm{yr}^{-1}\,\mathrm{kpc}^{-2}$ depending on the exact size of \tdor). C) Ratio of modelled to observed present-day mass-functions illustrating that the Salpeter IMF model underpredicts the number of massive stars in our sample, in particular above $30\,\msun$.}
\label{fig:sfh-imf}
\end{figure*}

We match the derived atmospheric parameters of the apparently single VFTS stars to stellar evolutionary models using the
Bayesian code \bonnsai, which has been successfully tested with high precision observations of 
Galactic eclipsing binary stars \cite{2014A&A...570A..66S}. 
\bonnsai takes uncertainties in the atmospheric parameters into account and determines full posterior 
probability distributions of stellar properties including the ages and initial masses of the VFTS stars \cite{Methods}. 
By summing these full posterior probability distributions 
of individual stars, we obtain the overall distributions of stellar ages and initial masses 
of massive stars currently present in \tdor (Fig.~\ref{fig:sfh-imf}). These distributions 
are missing those stars that already ended their nuclear burning. 
However, given that we know both the present-day 
age and mass distributions, we can correct for these missing stars
and derive the star-formation history (SFH) and IMF of massive stars in \tdor \cite{Methods}, 
allowing us to fully characterise this prototype starburst.

When determining the SFH and IMF, it is necessary to account for selection biases. 
The VFTS target selection implemented a magnitude cut, observing only stars brighter than 
17th magnitude in the \V-band \cite{2011A&A...530A.108E}. 
Compared to a full photometric census of massive stars in \tdor \cite{2013A&A...558A.134D}, 
the VFTS sample is about 73\% complete.
While the VFTS is incomplete for stars $\lesssim 15\,\msun$ because of the magnitude limit, 
the completeness shows no correlation with the \V-band magnitude of stars more massive than $15\,\msun$ \cite{Methods}. 
Of the 452 stars with robust stellar parameters, 247 are more massive than $15\,\msun$ and 
form the basis of our determination of the SFH and high-mass end of the IMF. 
Incompleteness corrections are applied to account for our selection process \cite{Methods}.
We assume the high-mass IMF is a power-law function, 
$\xi(M) \propto M^{-\gamma}$, where $M$ is the mass and $\gamma$ the slope,
and compute the SFH and corresponding prediction of the distribution of initial masses 
for different IMF slopes until we best match (i) the number of stars above a given mass 
and (ii) the observed initial-mass distribution \cite{Methods}.

% Number of massive stars for various IMF slopes
\begin{figure*}
\centering
\includegraphics[width=0.6\textwidth]{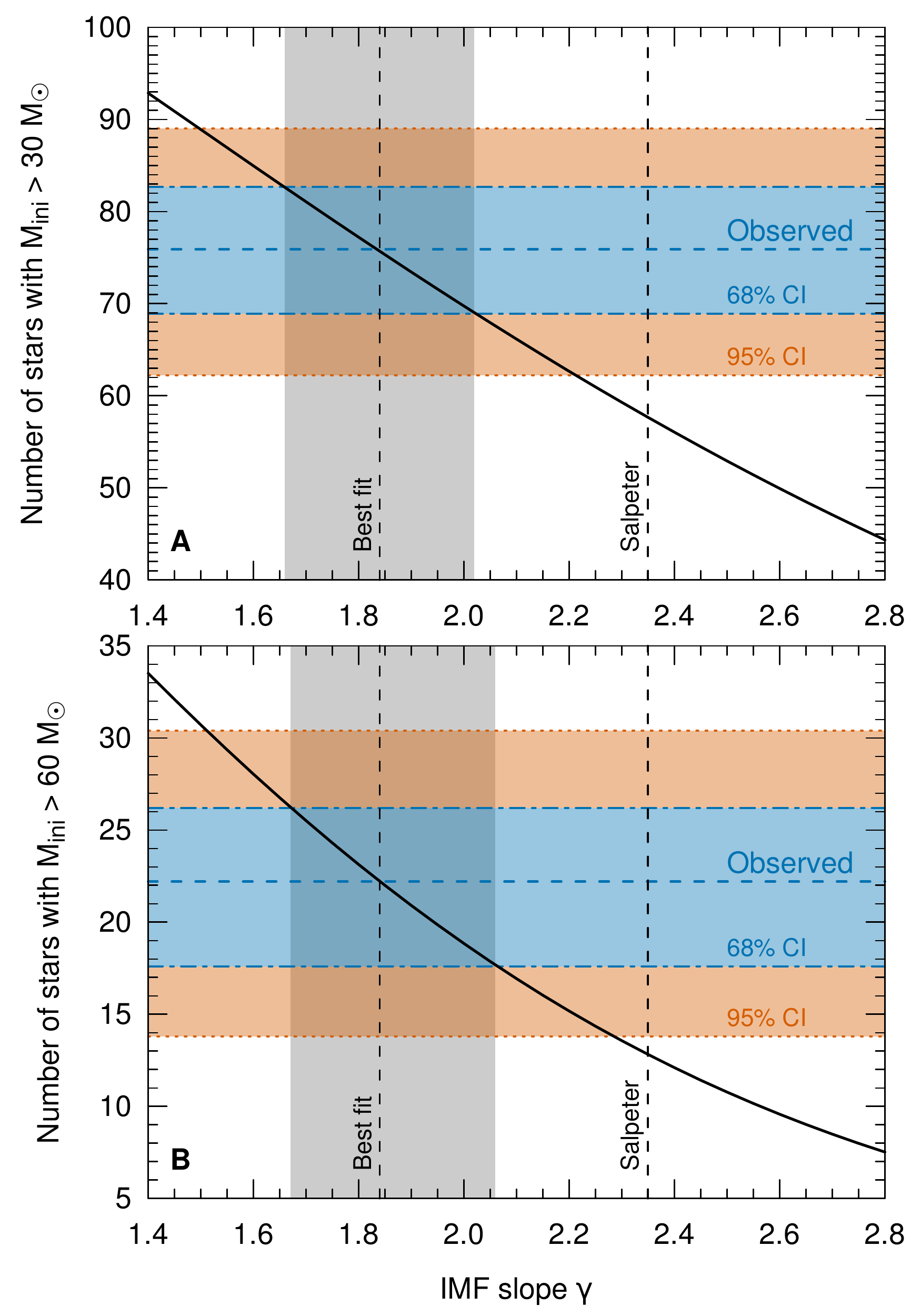}
\caption{\textbf{Expected number of massive stars in our sample initially more massive than (A) $30\,\msun$ and (B) $60\,\msun$ as a function of the IMF slope $\gamma$ (black solid line)}.
The blue and red shaded areas indicate the $68\%$ and $95\%$ confidence intervals of the observed number of stars, respectively (cf.\ Fig.~\ref{fig:number-of-stars-above-m}). The IMF slopes best reproducing the observed number of stars and the associated 68\% intervals are indicated by the vertical dashed lines and grey shaded regions and correspond to $\gamma=1.84^{+0.18}_{-0.18}$ and $\gamma=1.84^{+0.22}_{-0.17}$ for stars more massive than $30\,\msun$ and $60\,\msun$, respectively.}
\label{fig:number-stars-vs-gamma}
\end{figure*}

\begin{figure*}
\centering
\includegraphics[width=0.7\textwidth]{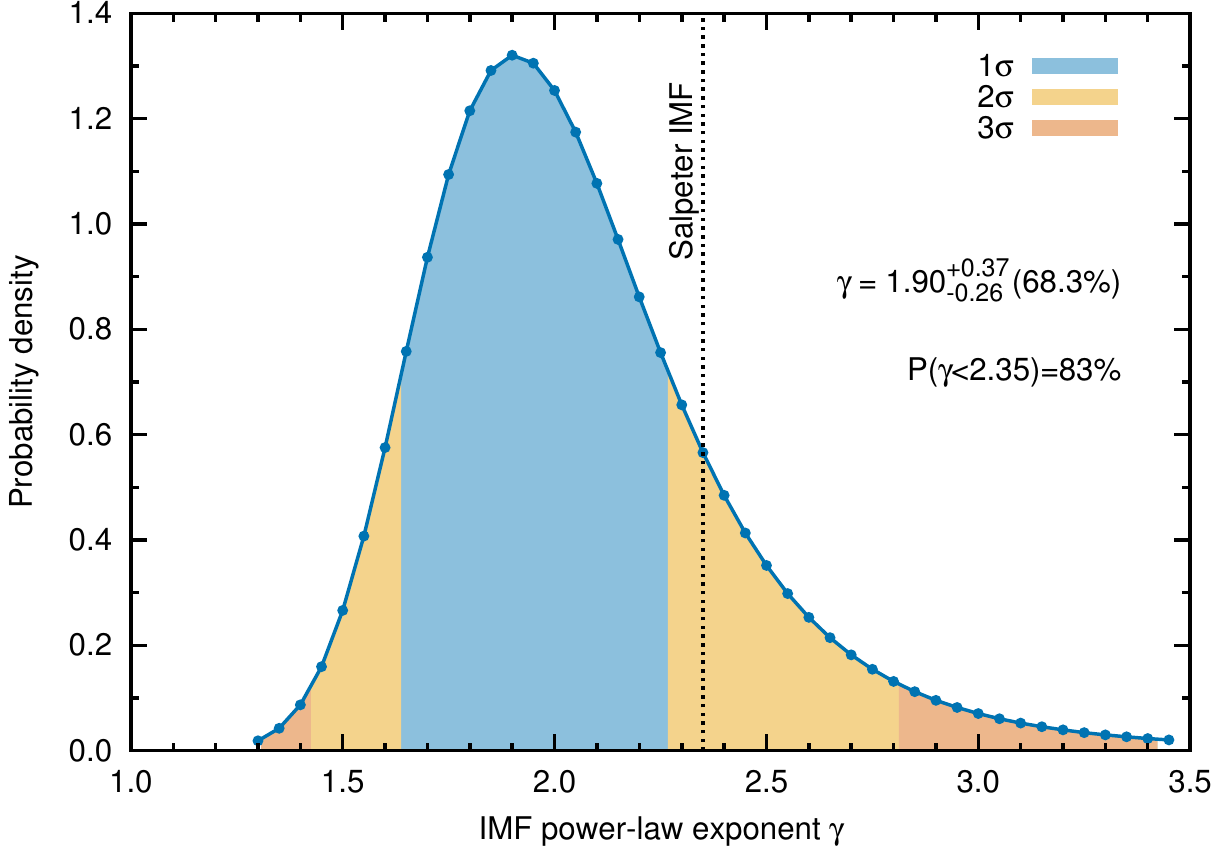}
\caption{\textbf{Probability density function of the inferred IMF slope in \tdor based on $\chi^2$ power-law fitting over the mass range $15\text{--}200\,\msun$.} 
The shaded areas represent $1\sigma$, $2\sigma$ and $3\sigma$ confidence regions and the slope of the Salpeter IMF is indicated by the vertical dashed line. Our inferred IMF is shallower than Salpeter ($\gamma=2.35$) with 83\% confidence.}
\label{fig:gamma}
\end{figure*}

We find that the observed distribution of initial masses of stars in \tdor 
is densely sampled up to about $200\,\msun$. It is shallower than that predicted by a Salpeter IMF
with $\gamma=2.35$ and the discrepancy increases with mass (Fig.~\ref{fig:sfh-imf}C).
Relative to Salpeter, we find an excess of $18.2^{+6.8}_{-7.0}$ 
($32^{+12}_{-12}\%$) stars more massive than $30\,\msun$
and $9.4^{+4.0}_{-4.6}$ ($73^{+31}_{-36}\%$) stars more massive than $60\,\msun$ (Figs.~\ref{fig:number-stars-vs-gamma} and~\ref{fig:number-of-stars-above-m};
unless stated otherwise, uncertainties are 68.3\% confidence intervals).
The hypothesis that a Salpeter IMF can explain the large number of stars more massive than $30\,\msun$
in our sample can thus be rejected with $>99\%$ confidence \cite{Methods}. 
The number of stars more massive than $30\,\msun$ are best reproduced by an 
IMF slope of $\gamma=1.84^{+0.18}_{-0.18}$ (Fig.~\ref{fig:number-stars-vs-gamma}).
Using our second diagnostic, a least-square fit to the observed distribution of initial masses
over the full mass range of $15\text{--}200\,\msun$, our best fit is $\gamma=1.90^{+0.37}_{-0.26}$
(Figs.~\ref{fig:sfh-imf} and~\ref{fig:gamma}), in agreement with our first 
estimate based on the number of massive stars $\geq 30\,\msun$.
Our high-mass IMF slope is shallower than the slope inferred for
stars below $\approx 20\,\msun$ in the vicinity of R136 by other studies \cite{2009ApJ...707.1347A,2015ApJ...811...76C}.

% PDF of IMF slope gamma

The limitation of our sample to stars $\geq 15\,\msun$ means that 
we can reconstruct the SFH of \tdor over the last $\approx 12\,\myr$.
When also considering the $1\text{--}2\,\mathrm{Myr}$ old stars in R136 
that were not observed within VFTS \cite{2016MNRAS.458..624C}, we find that the star-formation 
rate in \tdor sharply increased about $8\,\mathrm{Myr}$ ago and seems to have dropped 
about $1\,\mathrm{Myr}$ ago (Fig.~\ref{fig:sfh-imf}A).
If the currently observed drop continues for another Myr, the duration of the main star-forming event
will be shorter than about $10\,\mathrm{Myr}$. This result complements 
a recent study \cite{2015ApJ...811...76C}
which finds a similar time-dependence of star formation around the central R136 star cluster in \tdor based on 
photometric data of low- and intermediate-mass stars.
We therefore conclude that star formation in the \tdor starburst is synchronised across a wide mass range.

Our results challenge the suggested $150\,\msun$ limit \cite{2005Natur.434..192F} 
for the maximum birth mass of stars.
The most massive star in our sample, VFTS\,1025 (also known as R136c), 
has an initial mass of $203^{+40}_{-44}\,\msun$ \cite{Methods}.
From stochastic sampling experiments \cite{Methods}, we exclude maximum 
stellar birth masses of more than $500\,\msun$ in \tdor with 90\% confidence because 
we would otherwise expect to find at least one star above $250\,\msun$ in our sample. 
Our observations are thus consistent with the claim of stars with initial masses of up to $300\,\msun$ 
in the core of R136 \cite{2010MNRAS.408..731C}.

Approximately 15\%--40\% of our sample stars are expected to be 
products of mass transfer in binary star systems \cite{2014ApJ...782....7D}. 
Binary mass transfer in a stellar population produces a net surplus of massive stars and rejuvenates stars such that
they appear younger than they really are \cite{2014ApJ...780..117S}.
Mass accretion alone biases the inferred IMF slope to flatter values
whereas rejuvenation steepens it. Taken together, we calculate that 
these two effects roughly cancel out in our case and thus binary mass transfer
cannot explain the difference between our inferred IMF and that of Salpeter \cite{Methods}.
Also, our final sample of stars contains unrecognised binaries but
they do not affect our conclusions \cite{Methods}.

The core of the R136 star cluster is excluded from the VFTS, but stars ejected from R136 (so-called runaway stars)
may enter our sample. Runaway stars are biased towards high masses \cite{1961BAN....15..265B} and thus 
flatten the upper IMF. However, it is found that  
star clusters such as R136 typically eject about 5--10 stars above $15\,\msun$ 
\cite{2011Sci...334.1380F,2015ApJ...805...92O} which
is insufficient to explain the expected excess of $25\text{--}50$ stars 
above $30\,\msun$ in \tdor, after correcting for the completeness of our 
sample and that of the VFTS \cite{Methods}.

We conclude that the \tdor starburst has
produced stars up to very high masses ($\gtrsim 200\,\msun$), with a statistically significant
excess of stars above $30\,\msun$ and an IMF shallower above $15\,\msun$
than a Salpeter IMF.
Measuring the IMF slope above $30\text{--}60\,\msun$
has proven difficult \cite{2010ARA&A..48..339B} and in general
large uncertainties in the high-mass IMF slope remain \cite{2013ApJ...762..123W}.
This raises the question of whether star formation in \tdor proceeded differently.
It has been suggested that starburst regions themselves
provide conditions for forming relatively more massive stars by 
the heating of natal clouds from nearby and previous generations of stars \cite{2005MNRAS.359..211L}.
Alternatively, a lower metallicity may lead to the formation of more massive stars 
because of weaker gas cooling during star formation. 
An IMF slope shallower than Salpeter may then be expected 
at high redshift when the Universe was hotter and the metallicity lower \cite{1999ApJ...527L...5B,2005MNRAS.359..211L}. %2002Sci...295...93A

Because massive-star feedback increases steeply with stellar mass, it is strongly affected by the IMF slope.
Comparing an IMF slope of $\gamma=1.90^{+0.37}_{-0.26}$ to Salpeter,
we expect $70^{+10}_{-60}\%$ more core-collapse supernovae and an increase of 
supernova metal-yields and hydrogen ionising radiation 
by factors of $3.0^{+1.6}_{-1.8}$ and $3.7^{+2.4}_{-2.4}$, respectively \cite{Methods}.
The formation rate of black holes increases by a factor of $2.8^{+1.0}_{-1.6}$ \cite{Methods}, directly affecting
the expected rate of black hole mergers found through their gravitational wave signals.
We also expect an increase in the predicted number of exotic transients that are
preferentially found in starbursting, metal-poor dwarf galaxies such as long duration gamma-ray bursts \cite{2017ApJ...834..170G}
and hydrogen-poor superluminous supernovae \cite{2016A&A...593A.115J}.
Many population synthesis models and large-scale cosmological simulations
assume an IMF that is truncated at $100\,\msun$. 
Compared to those, the various factors estimated above are even larger \cite{Methods}.

% moved ack from original position higher up (before refs)
\paragraph*{Acknowledgements}
We thank the referees for constructive feedback that helped improve this work. Based on observations collected at the European Southern Observatory under program ID 182.D-0222. This work was supported by the Oxford Hintze Centre for Astrophysical Surveys which is funded through generous support from the Hintze Family Charitable Foundation. HS acknowledges support from the FWO-Odysseus program under project G0F8H6N. GG acknowledges financial support from the Deutsche Forschungsgemeinschaft, Grant No.\ GR 1717/5. OHRA acknowledges funding from the European Union's Horizon 2020 research and innovation programme under the Marie Sk{\l}odowska-Curie grant agreement No 665593 awarded to the Science and Technology Facilities Council. CS-S acknowledges support from CONICYT-Chile through the FONDECYT Postdoctoral Project No.~3170778. SSD and AH thank the Spanish MINECO for grants AYA2015-68012-C2-1 and SEV2015-0548. SdM has received funding under the European Union’s Horizon 2020 research and innovation programme from the European Commission under the Marie Sk{\l}odowska-Curie (Grant Agreement No. 661502) and the European Research Council (ERC, Grant agreement No. 715063). MGa and FN acknowledge Spanish MINECO grants FIS2012-39162-C06-01 and ESP2015-65597-C4-1-R. MGi acknowledges financial support from the Royal Society (University Research Fellowship) and the European Research Council (ERC StG-335936, CLUSTERS). RGI thanks the STFC for funding his Rutherford fellowship under grant ST/L003910/1 and Churchill College, Cambridge for his fellowship and access to their library. VK acknowledges funding from the FONDECYT-Chile fellowship grant No.\ 3160117. JMA acknowledges support from the Spanish Government Ministerio de Econom{\'i}a y Competitividad (MINECO) through grant AYA2016-75 931-C2-2-P. NM acknowledges the financial support of the Bulgarian NSF under grant DN08/1/13.12.2016. STScI is operated by AURA, Inc. under NASA contract NAS5-26555. CJE is also a Visiting Professor at the University of Edinburgh. The raw VFTS observations are available from the European Southern Observatory's Science Archive Facility at \href{http://archive.eso.org}{http://archive.eso.org} under project ID 182.D-0222. Tabulated data for the input and derived stellar parameters used in this study, the best-fitting SFH, and our Python code for determining the stellar maximum birth masses are all provided in the supplementary material. A web interface for the \bonnsai software is available at \href{http://www.astro.uni-bonn.de/stars/bonnsai}{http://www.astro.uni-bonn.de/stars/bonnsai}.

\renewcommand{\refname}{References and Notes}
\bibliographystyle{Science}
%\bibliography{bibliography,additional_bibliography}

\end{multicols}

% Supplementary Material
\clearpage

% Single-space the supplement
\baselineskip12pt

% reset different counters
\setcounter{page}{1}
\setcounter{figure}{0}
\setcounter{table}{0}
% add an "S" to labels
\renewcommand{\thefigure}{S\arabic{figure}}
\renewcommand{\thetable}{S\arabic{table}}
\renewcommand{\thesection}{S\arabic{section}}
\renewcommand{\theequation}{S.\arabic{equation}}

% design front page as required by Science; see their Word template at
% www.sciencemag.org/site/feature/contribinfo/prep/Science_Supplementary_Materials_Word_template.doc

\begin{center}
\includegraphics[width=4.13cm]{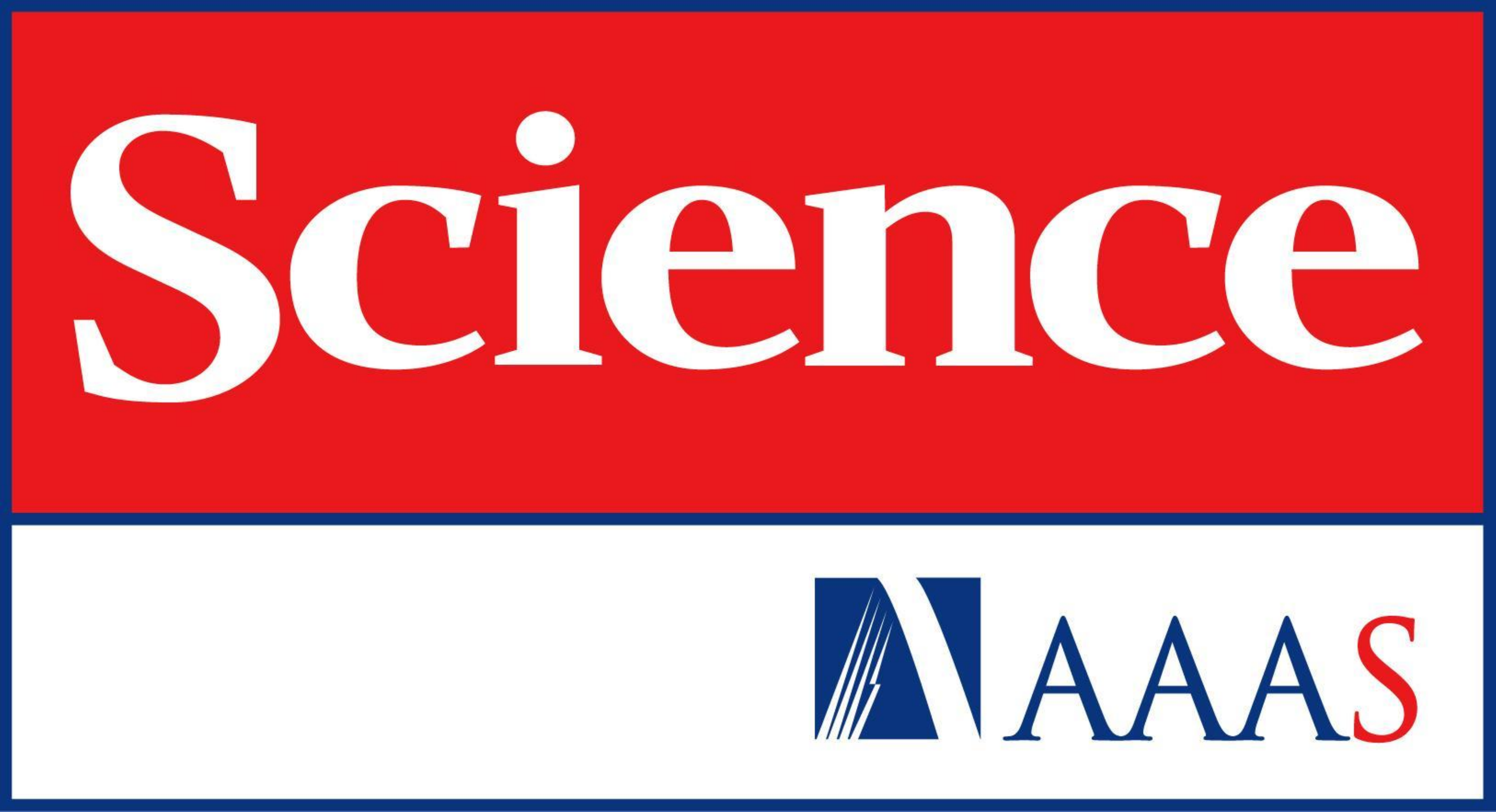}
\end{center}

\begin{center}
\LARGE{Supplementary Materials for}
\end{center}

\begin{center}
\large{\mytitle}
\end{center}

\begin{center}
F.R.N.~Schneider, H.~Sana, C.J.~Evans, J.M.~Bestenlehner, N.~Castro, L.~Fossati, G.~Gr{\"a}fener, N.~Langer, O.H.~Ram{\'i}rez-Agudelo,~C. Sab{\'i}n-Sanjul{\'i}an, S.~Sim{\'o}n-D{\'i}az, F~Tramper, P.A.~Crowther, A.~de Koter, S.E.~de Mink, P.L.~Dufton, M.~Garcia, M.~Gieles, V.~H\'{e}nault-Brunet, A.~Herrero, R.G.~Izzard, V.~Kalari, D.J.~Lennon, J.~Ma\'{i}z~Apell\'{a}niz, N.~Markova, F.~Najarro, Ph.~Podsiadlowski, J.~Puls, W.D~Taylor, J.Th.~van~Loon, J.S.~Vink, and C.~Norman
\end{center}

\begin{center}
correspondence to: \href{mailto:fabian.schneider@physics.ox.ac.uk}{fabian.schneider@physics.ox.ac.uk}
\end{center}

\vspace{0.5cm}

\noindent
\textbf{This PDF file includes:}
\vspace{0.3cm}

Materials and Methods

Supplementary Text

Table S1 to S3

Fig S1 to S10

\vspace{0.5cm}

\noindent
\textbf{Other Suplementary Materials for this manuscript includes the following:}
\vspace{0.3cm}

Machine-readable version of best-fitting SFH

Python program used for constraining limits on the maximum birth mass of stars

Machine-readable table of stellar parameters (Table~S3)

\clearpage
\setlength{\parindent}{4ex}
\setlength{\parskip}{1.2mm}

\section*{Materials and Methods}

Deriving the SFH and IMF of massive stars in \tdor from spectroscopic observations invokes a number of steps and techniques. To aid the understanding of our approach, we first give a simplified overview of the whole procedure (Sect.~\ref{sec:method-summary}) before discussing the individual steps in more detail (Sects.~\ref{sec:sample-selection}--\ref{sec:imf-sfh}).

\section{Method overview}\label{sec:method-summary}

The starting point of our investigation is the collection of spectra of more than 800 massive stars in \tdor observed within the VFTS \cite{2011A&A...530A.108E}. After identifying spectroscopic binaries and visual multiples, the spectra of the individual objects have been modelled with state-of-the-art atmosphere codes to obtain parameters such as effective temperature and surface gravity (Sect.~\ref{sec:atmospheric-parameters}). For example, the effective temperature of VFTS\,249 is found to be $T_\mathrm{eff}=36500{\pm}760\,\mathrm{K}$, the luminosity $\log L/\lsun=4.78{\pm}0.14$, the surface gravity $\log g/\mathrm{cm}\,\mathrm{s}^{-2}=4.11{\pm}0.11$ and the projected rotational velocity $v\sin i=300{\pm}30\,\kms$ \cite{2014A&A...564A..39S,2017A&A...601A..79S}.

\begin{figure}
\centering
\includegraphics[width=0.57\textwidth]{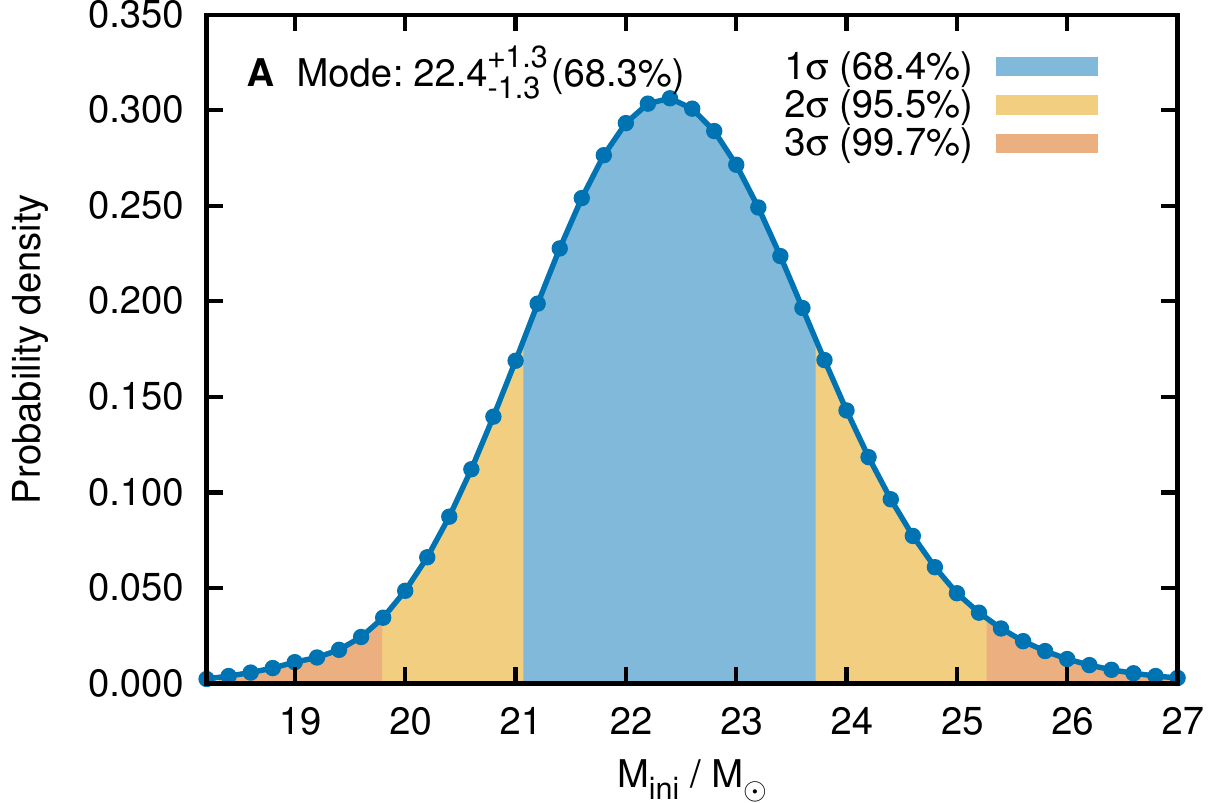}
\includegraphics[width=0.57\textwidth]{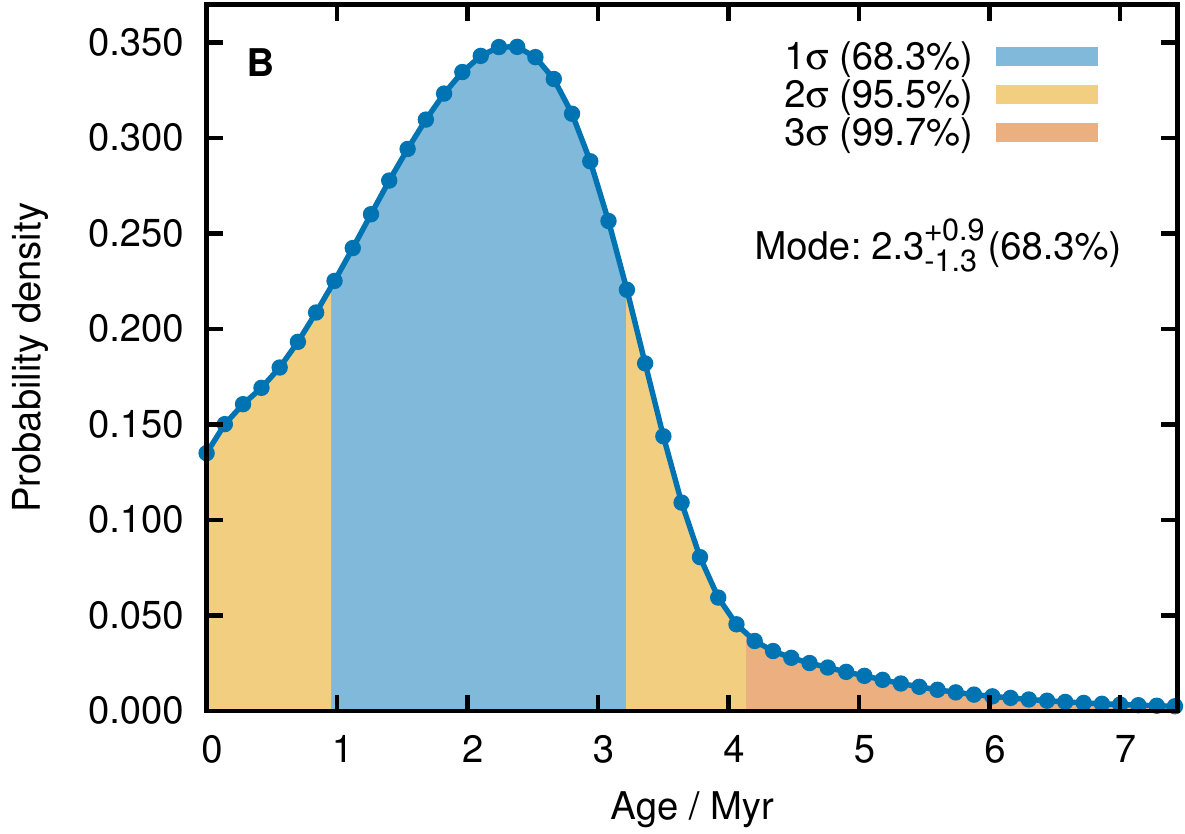}
\includegraphics[width=0.57\textwidth]{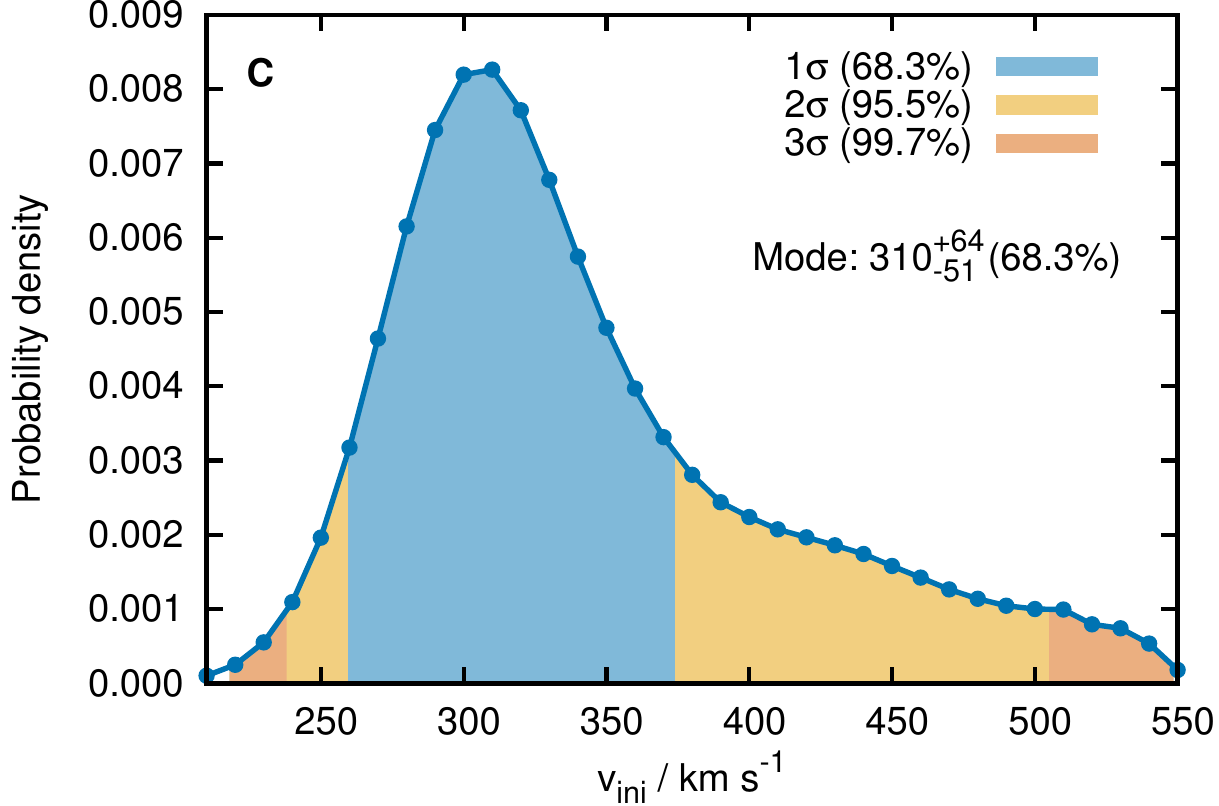}
\caption{\textbf{Posterior probability-density functions of the inferred stellar model parameters initial mass (A), age (B) and initial rotational velocity (C) of VFTS\,249.} The shaded areas are the $1\sigma$, $2\sigma$ and $3\sigma$ confidence regions, and summary statistics, \ie the mode values of the distributions and the corresponding $1\sigma$ confidence levels, are provided.}
\label{fig:pdf-examples}
\end{figure}

We then match the determined atmospheric parameters against rotating, single-star models \cite{2011A&A...530A.115B,2015A&A...573A..71K} using the Bayesian code \bonnsai \cite{2014A&A...570A..66S,2017A&A...598A..60S,Bonnsai}. Because this is a Bayesian framework, we take uncertainties in the atmospheric parameters and prior knowledge fully into account and obtain posterior probability distributions of the model parameters initial mass, age and initial rotational velocity (Sect.~\ref{sec:stellar-parameters}). For our example star VFTS\,249, these distributions are shown in Fig.~\ref{fig:pdf-examples}. The distributions are not necessarily Gaussian and are usually asymmetric and they allow us to define summary statistics (mode values including 68.3\% confidence intervals) for initial mass, age and initial rotational velocity of $22.4^{+1.3}_{-1.3}\,\msun$, $2.3^{+0.9}_{-1.3}\,\myr$ and $310^{+64}_{-51}\kms$, respectively.

Robust atmosphere modelling is not always possible because of composite spectra, nebular contamination, binarity and contamination from bright stars in nearby fibres (Sect.~\ref{sec:sample-selection}). Also, the stellar models are unable to reproduce the atmospheric parameters of some stars within their uncertainties. In total, we are able to determine full posterior probability distributions of initial mass, age and other stellar parameters from robust atmospheric parameters for 452 VFTS stars. Summary statistics of our full posterior probability distributions together with the atmospheric parameters used to infer them are provided in Table~\ref{tab:stellar-parameters} for all 452 targets.

Because of the magnitude limit and target selection of the VFTS, there are no biases regarding the completeness of stars initially more massive than $15\,\msun$ such that our VFTS sample can be regarded as representative of the massive \tdor stellar population (Sect.~\ref{sec:sample-selection}). This is essential when deriving the SFH and IMF, so we continue our work with the 247 VFTS stars that are found to be initially more massive than $15\,\msun$. For these stars, we take the full posterior probability distributions of initial mass and age (cf. Figs.~\ref{fig:pdf-examples}A and~\ref{fig:pdf-examples}B) and sum them, resulting in posterior density functions of initial mass and age of 247 stars with initial masses $\geq 15\,\msun$ in \tdor. When adding the individual contributions of our VFTS targets, we correct for the selection process (Sect.~\ref{sec:inferring-age-mass-distributions}). We finally use a bootstrapping method to estimate uncertainties in the obtained distributions of initial mass and age. The final distributions and their uncertainties are shown in Fig.~\ref{fig:sfh-imf} (solid black lines and blue shaded regions, respectively).

The obtained probability distributions of initial mass and age of our sample of 247 VFTS stars are neither IMFs nor SFHs because they lack those stars that already ended nuclear-burning. Stars in \tdor are not coeval and we can also not assume that the star-formation rate was constant in the past. This makes the inference of the IMF and SFH more interdependent. To account for this, we have developed an iterative process to simultaneously infer the IMF and SFH from our combined distributions of initial mass and age of the 247 VFTS targets. This method is described in Sect.~\ref{sec:imf-sfh} and yields the best-fitting IMF and SFH of massive stars ($\geq 15\,\msun$) in \tdor shown as the red curves in Fig.~\ref{fig:sfh-imf} and the full probability distribution of the inferred IMF slope (Fig.~\ref{fig:gamma}).

\section{Sample selection}\label{sec:sample-selection}

When deriving the SFH and IMF of any stellar population, it is crucial that the sample of stars is observationally unbiased and as complete as possible (\ie representative of the whole stellar population). All selection criteria must be understood and properly accounted for. The only selection criteria of VFTS targets in \tdor are that (i) the stars are brighter than a \V-band magnitude of $V=17\,\mathrm{mag}$ and (ii) as many targets as possible can be observed with the \flames fibre set-ups \cite{2011A&A...530A.108E}. This means that neither bright nor dim targets have been preferentially selected and that crowded regions such as the core of the R136 star cluster have been avoided because of the $1.2\,\mathrm{arcsec}$ size of each of the \flames fibres on the sky. Except for VFTS\,1025, there are no stars in our sample closer than $\approx0.2\text{--}0.3\,\mathrm{arcmin}$ to R136, corresponding to about $3\text{--}4\,\mathrm{pc}$ at a distance of $50\,\mathrm{kpc}$ to \tdor \cite{2013Natur.495...76P}.

\begin{figure}[t]
\centering
\includegraphics[width=0.8\textwidth]{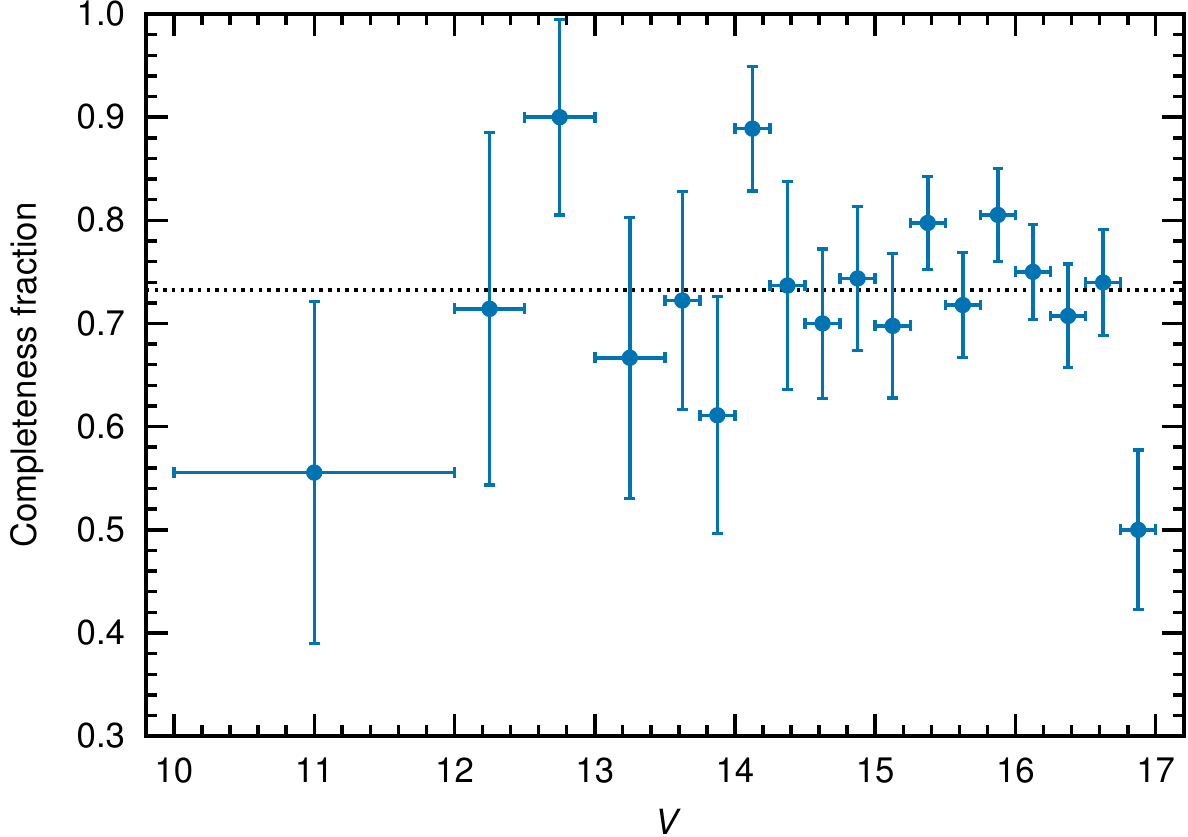}
\caption{\textbf{Completeness fraction of the VFTS sample as a function of \V-band magnitude.} The completeness is with respect to a full census of hot and luminous stars in \tdor \cite{2013A&A...558A.134D} and is on average 73\% as indicated by the black dotted line. The error bars in \V-band magnitude indicate the bin-widths used to compute the completeness fraction.}
\label{fig:vfts-completeness}
\end{figure}

To probe whether there are nevertheless hidden biases in the VFTS sample, we compute the completeness of VFTS stars as a function of \V-band magnitude relative to a census of hot and luminous stars in \tdor (from data in fig.~6 of Ref.~\citenum{2013A&A...558A.134D}). The VFTS completeness fraction is constant over the whole \V-band magnitude range and on average about 73\% (Fig.~\ref{fig:vfts-completeness}). The completeness only drops around the $V=17\,\mathrm{mag}$ threshold. In what follows, we only consider stars more massive than $15\,\msun$ such that this drop does not affect our work because these stars have $V<16.5\,\mathrm{mag}$ given the distance to \tdor and its reddening conditions. The VFTS sample might be slightly less complete at the high luminosity end ($V=10\text{--}12\,\mathrm{mag}$) but, given the low number of stars and hence high Poisson uncertainty in the completeness at these bright magnitudes, this offset does not seem to be significant. If it were, we would underestimate the number of very massive stars, which would only strengthen our conclusions.

% VFTS completeness

Modelling composite spectra is more difficult than single star spectra and may result in more uncertain stellar parameters and systematic biases, especially if composite spectra are treated as originating from only one source. In the VFTS, composite spectra arise whenever more than one star contributes noticeably to the light in one of the fibres of the \flames instrument used for observations. The multi-epoch nature of the VFTS allows for the identification of spectroscopic binaries \cite{2013A&A...550A.107S,2015A&A...580A..93D}, and visual multiples could be identified by comparing the position and sizes of the fibres on the sky with high-resolution Hubble Space Telescope images \cite{2014A&A...564A..40W,2015A&A...580A..93D}. Binary stars and higher-order multiple systems, visual multiples, nearby bright stars and other contaminating sources can potentially produce composite spectra. In order to minimise potential biases and utilise only robust stellar parameters, we disregard all known spectroscopic binaries (231 stars) and visual multiples (58 stars; 14 stars are both visual multiples and spectroscopic binaries).

Among the remaining 626 stars, satisfactory spectral fits could not always be achieved, \eg because of insufficient data quality, mostly low signal-to-noise for stars close to the $V=17\,\mathrm{mag}$ limit or stars suffering from heavy nebular contamination (Sect.~\ref{sec:atmospheric-parameters}). Furthermore, the evolutionary models we use \cite{2011A&A...530A.115B,2015A&A...573A..71K} are not always able to reproduce the derived atmospheric parameters (Sect.~\ref{sec:stellar-parameters}). We disregard such stars from further analysis (37 and 35 stars, respectively). 

Stars cooler than $9000\,\mathrm{K}$ have also been removed from our final sample (88 stars) because the stellar models do not cover this evolutionary phase and it is difficult to obtain good ages and masses for these stars. Importantly, there are only a few stars (at most 4 of the 88 stars) that might be more massive than $15\,\msun$ and younger than $10\,\myr$. Their exclusion does therefore not influence our results noticeably.

Our full sample of stars with robust fundamental parameters consists of 452 apparently single VFTS stars outside dense cluster cores. In terms of spectral types these are 13 WNh and Of/WN, 4 classical WR, 173 O-type, 258 B-type and 4 A-type stars. This sub-sample of VFTS stars is no longer fully representative of the \tdor massive star population because we, \eg, remove stars with composite spectra. In Sect.~\ref{sec:inferring-age-mass-distributions}, we describe how we correct for these selection effects.

\section{Atmospheric parameter determination}\label{sec:atmospheric-parameters}

The atmospheric analysis of the VFTS stars has been performed over several years by the VFTS consortium. We briefly summarise the corresponding sources and provide details of new atmospheric analyses. The uncertainties on the determined stellar parameters are often only statistical errors, and we therefore apply typical minimum $1\sigma$ uncertainties of $500\,\mathrm{K}$ in effective temperature, $0.1\,\mathrm{dex}$ in logarithmic luminosity, $0.1\,\mathrm{dex}$ in logarithmic surface gravity and 10\% or at least $30\,\kms$ in projected rotational velocity if the atmosphere analyses provide smaller uncertainties. Different atmosphere analysis codes were applied and consistency checks were carried out to ensure that the different approaches give comparable results (\eg Sect.~3.4 of \cite{2017A&A...600A..81R}).

\subsection{Wolf--Rayet and slash stars}\label{sec:pd-wr-stars}

Based on their spectral morphology the Wolf--Rayet (WR) stars in the VFTS have been divided into one group containing hydrogen-rich Of/WN (hereafter called slash stars) and WNh stars that are most likely still in the phase of core hydrogen burning, and a second group containing evolved WN and WC stars in the phase of core helium burning. For the second group, we use the atmospheric parameters of \cite{2013A&A...558A.134D} determined by photometric calibrations and surface abundances from \cite{2002A&A...392..653C,2014A&A...565A..27H}. The analysis of the slash and WNh stars is taken from \cite{2014A&A...570A..38B}.

\subsection{O and B stars}\label{sec:pd-ob-stars}

The stellar parameters of the O stars have been determined by modelling the VFTS spectra with \fastwind \cite{1997A&A...323..488S,2005A&A...435..669P,2012A&A...537A..79R}. The stellar atmosphere code \fastwind provides synthetic spectra of O- and B-type stars taking non-local thermodynamic equilibrium effects in spherical symmetry with an explicit treatment of the stellar wind into account. The resulting O-star atmospheric parameters adopted here have been previously published in two samples separated by their luminosity class: giants and supergiants \cite{2017A&A...600A..81R}, and dwarfs and sub-giants \cite{2014A&A...564A..39S,2017A&A...601A..79S}. Atmospheric parameters for the B-type supergiants were derived by \cite{2015A&A...575A..70M}. 

The spectra of the remaining B stars are modelled following the $\chi^2$ fitting technique described in \cite{2012A&A...542A..79C} (see also \cite{2007PhDT.......308L,2010A&A...515A..74L,2011JPhCS.328a2021S}). The algorithm uses a pre-computed \fastwind stellar atmosphere grid. Nebular emission in the spectra is manually trimmed out, avoiding contamination in the quantitative analyses.

The available atmosphere model grid was computed at solar metallicity and covers effective temperatures of $12,000\text{--}34,000\,\mathrm{K}$ and surface gravities $\log g$ of $2.0\text{--}4.4\,\mathrm{dex}$ in steps of $1000\,\mathrm{K}$ and $0.1\,\mathrm{dex}$, respectively. The different metallicity between the grid and the VFTS stars (about 40\% solar) affects the derived effective temperatures and hence our estimates of ages and masses because of differences in the effects of line blanketing. We explore this bias in a few test cases for which we have atmosphere models with the appropriate metallicity of the Large Magellanic Cloud (LMC) and find that our temperature determinations are on average too cool by about $2000\,\mathrm{K}$ for stars with $T_\mathrm{eff}<25,000\,\mathrm{K}$ and too hot by about $1000\,\mathrm{K}$ for $T_\mathrm{eff}>25,000\,\mathrm{K}$. The surface gravities change correspondingly by about $\pm0.1\,\mathrm{dex}$. Depending on the exact temperatures, gravities and luminosities, these biases influence the inferred ages and masses at a 10\%-20\% level. Temperatures which are cooler in the model than in reality result in older and less massive stars, and vice versa.

As expected given their spectral type, we find that most of the B-stars analysed in this way are initially less massive than $15\,\msun$ and older than $8\text{--}10\,\mathrm{Myr}$. Our final sample only contains stars more massive than $15\,\msun$ such that our results and conclusions are essentially unaffected. Less than 10\% of all VFTS B-dwarfs end up in the final sample because of the $15\,\msun$ mass cut (Sect.~\ref{sec:discussion-b-stars} for more details).

\subsection{Stars of A-type and later}\label{sec:pd-cool-stars}

For the cooler stars (A-type and later) we adopt effective temperatures on the basis of their spectral types, interpolating between the Galactic and Small-Magellanic-Cloud values for A-, F- and G-type stars \cite{2003MNRAS.345.1223E}. For the mid-late K and early M stars, we adopt $T_\mathrm{eff}=4100{\pm}150\,\mathrm{K}$ and $4000{\pm}150\,\mathrm{K}$, respectively \cite{2013ApJ...767....3D,2015ApJ...806...21D}. Between these regimes---i.e. stars classified as ``late G/early K''---we adopt $T_\mathrm{eff}=4750{\pm}650\,\mathrm{K}$, which is the approximate mid-point between the interpolated value for G5 (5375\,K) and the 4100\,K for the late K-type stars (with a large uncertainty given the assumptions/interpolations).

We determine bolometric luminosities using K-band photometry from the near-infrared $YJK_\mathrm{s}$ VISual-and-Infrared-Telescope-for-Astronomy survey of the Magellanic Clouds \cite{2011A&A...527A.116C} and The Two Micron All Sky Survey \cite{2006AJ....131.1163S} if needed, and use bolometric $K$-band corrections over the effective temperature range $10,000\text{--}4,000\,\mathrm{K}$ with a half-solar metallicity and surface gravity $\log g=2.0$ \cite{2006A&A...450..735M}. The adopted bolometric corrections are extrapolated to cool temperatures ($<5500\,\mathrm{K}$), in good agreement with other results \cite{2010MNRAS.403.1592B}. We use an average K-band extinction $A_\mathrm{K}=0.2\,\mathrm{mag}$ \cite{2013A&A...554A..33T,2017A&A...600A..81R}. Of these later-type stars, only one, VFTS\,820, ends up in our final sample (see Sect.~\ref{sec:stellar-parameters}). The other A-star in our final sample, VFTS\,739, has been analysed with the methods described in Sect.~\ref{sec:pd-ob-stars}.

\section{Stellar parameter determination}\label{sec:stellar-parameters}

The majority of our stars are in their main-sequence (MS) phase. They are thus covered by the single-star models of \cite{2011A&A...530A.115B,2015A&A...573A..71K} such that we can use the Bayesian code \bonnsai \cite{2014A&A...570A..66S,2017A&A...598A..60S} to determine, for each star, full posterior probability distributions of fundamental stellar parameters such as mass and age (rotating, single-star models that also cover the post-MS phase are currently not implemented in \bonnsai). To that end, we simultaneously match all available observables (in most cases effective temperature, surface gravity, luminosity and projected rotational velocity) to the stellar models while taking observed uncertainties and prior knowledge into account. We assume that all initial masses and ages are a priori equally probable. In principle, a Salpeter initial mass function \cite{1955ApJ...121..161S} and the observed star-formation history of \cite{2015ApJ...811...76C} for stars in NGC~2070 could have been used as prior distributions for initial mass and age, respectively, but we wish to derive mass and age distributions of our sample stars independently of such prior knowledge to probe mass functions and star formation in \tdor without introducing possible biases. As a prior distribution of initial rotational velocities, we use the observed distributions of rotational velocities of the apparently single VFTS O \cite{2013A&A...560A..29R} and B stars \cite{2013A&A...550A.109D}. We further assume that all rotation axes are randomly oriented in space when computing projected rotational velocities.

\bonnsai allows us to test whether the derived atmospheric parameters of stars can be reproduced by the stellar models. To that end, \bonnsai conducts a Pearson's $\chi^2$-hypothesis test and posterior predictive checks that take the full posterior probability distribution into account to determine whether the predictions of the stellar models--given the determined model parameters--are in agreement with observations \cite{2014A&A...570A..66S}. In both tests, we apply a significance level of $5\%$, \ie if one or both tests fail we are confident at $\geq 95\%$ that the stellar models are unable to reproduce all observables simultaneously within the observed uncertainties. Stars for which those tests fail are excluded from further analysis (about 7\% of all considered stars; see Table~\ref{tab:sample-summary-stats} below).

Post main-sequence stars and classical WR stars are not covered by the single-star models we employ, so we use alternative techniques to derive ages and masses for them. Stars in the Hertzsprung gap (HG) between the main-sequence and red supergiant phase evolve at nearly constant luminosity. The masses of such HG stars can therefore be inferred from the masses of stars at the terminal-age main-sequence. Because the luminosity is only approximately constant for stars crossing the HG gap, we increase the luminosity uncertainty by a factor of 2 when matching the derived luminosities to our terminal-age main-sequence, single-star models. The ages derived in this way correspond to the MS lifetimes of stars but the HG stars must be older than that, providing a lower age limit. After finishing core hydrogen burning, stars undergo nuclear burning for another $\approx 10\%$ of the MS lifetime before they end their nuclear burning lifetime. To be conservative, we decrease and increase the lower and upper age limits, respectively, by 10\%. The age probability distribution is then assumed to be uniform between these lower and upper age estimates.

Our age determination for the four classical WR stars is based on rotating evolutionary tracks of LMC metallicity \cite{2005A&A...429..581M}. Their most massive models enter the WR stage at an age of about $2.5\,\myr$, which can be regarded as a lower limit for the age of evolved WR stars. To estimate a conservative upper age limit, we consider Galactic models \cite{2003A&A...404..975M} and use the $8.5\,\myr$ lifetime of an initially $25\,\msun$ star as the maximum WR lifetime. To refine these rough age limits, we estimate the times at which different evolutionary tracks display surface compositions in agreement with the observed spectral types and hydrogen surface mass fractions. As stars of different mass enter and leave the respective phases at different luminosities, we can use the observed luminosities to constrain the ages of the sample stars. The accuracy of this approach is chiefly determined by the uncertainties on the luminosities. We adopt $\pm0.1\,\mathrm{dex}$ for logarithmic luminosities that are derived from spectrophotometric data and $\pm0.2\,\mathrm{dex}$ for logarithmic luminosities that are based on a combination of spectral synthesis and photometry. The initial masses are determined analogously to the ages, by interpolating between the evolutionary tracks that match the observed luminosities. Finally, present-day masses are derived from the mass--luminosity relation of core helium-burning stars \cite{2011A&A...535A..56G}, and the obtained age and mass limits are converted into uniform probability distributions bounded by the limits.

\begin{figure}[p]
\centering
\includegraphics[width=0.93\textwidth]{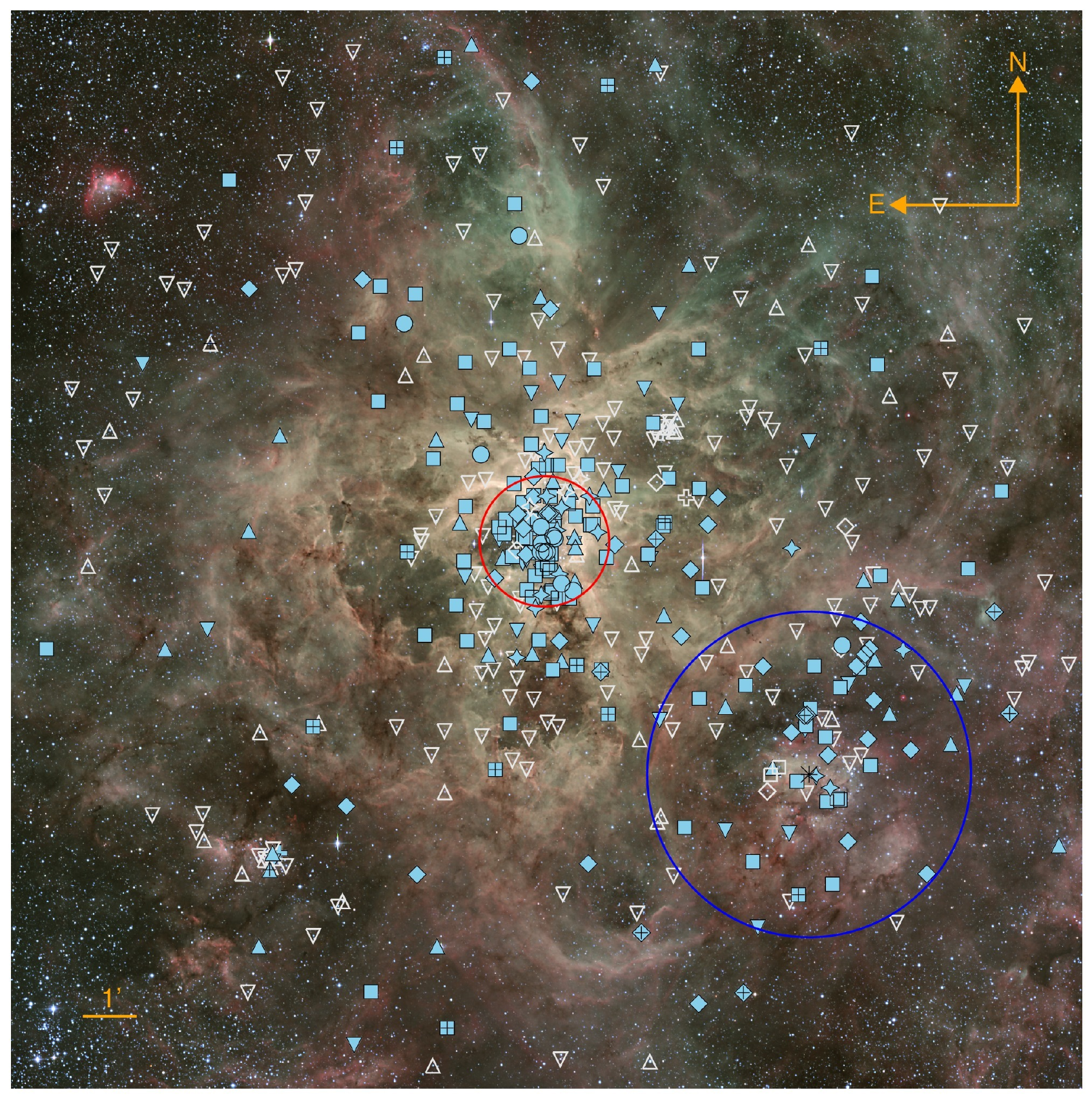}
\caption{\textbf{Positions of our sample stars in \tdor.} Open symbols indicate all 452 stars in our full sample, and the filled symbols those stars that are more massive than $15\,\msun$ and are used to derive the SFH and IMF of massive stars in \tdor. Circles denote slash/WNh/WR stars, squares O dwarfs, diamonds O giants, star symbols O-type stars without luminosity class, upward triangles B giants, downward triangles B dwarfs and pluses later-type stars. We further mark runaway candidates \cite{2014A&A...564A..40W,2015A&A...574A..13E} by additional plus signs and the position of the pulsar \pulsar by a black asterisk. The red and blue circles indicate the NGC~2070 (including R136) and NGC~2060 regions, respectively. The figure is centred on the R136 star cluster (RA 05h 38m 42.396s and Dec -69$^\circ$ 06' 03.36''). At a distance of $50\,\mathrm{kpc}$ to \tdor \cite{2013Natur.495...76P}, the one arcminute scale bar shown corresponds to $14.6\,\mathrm{pc}$. The background image is based on observations made with ESO Telescopes at the La Silla Observatory under programme ID 076.C-0888, processed and released by the ESO VOS/ADP group \cite{30Dor}.}
\label{fig:30dor}
\end{figure}

All observables and derived stellar parameters for our sample stars are summarised in Table~\ref{tab:stellar-parameters}, including a flag indicating which methods have been used for the determination of the atmospheric parameters. A summary of the number of stars in VFTS and in our sample is provided in Table~\ref{tab:sample-summary-stats} and the positions of all analysed VFTS stars in \tdor are illustrated in Fig.~\ref{fig:30dor}.

% summary statistics of stars in VFTS
\begin{table}[t]
\caption{\label{tab:sample-summary-stats}Summary statistics for all stars in the VFTS, our full sample for which we provide stellar parameters and our final sample of stars more massive than $15\,\msun$ used to constrain the SFH and IMF of \tdor. O-type stars without luminosity class (LC) are denoted "O no LC". Stars not reproduced refer to cases where the stellar evolution models cannot reproduce all observables simultaneously within the uncertainties and the column ``Not reproduced'' lists those stars. The ``Discarded'' column contains objects with composite spectra, uncertain atmospheric parameters etc. Most of the discarded, hot (earlier than A-type) stars are spectroscopic binaries.}
\centering
\begin{tabular}{cccccc}
\toprule 
 & VFTS & Full sample & Final sample & Not reproduced & Discarded\\
 & \cite{2011A&A...530A.108E} & (this work) & (this work) & (this work) & (this work)\\
\midrule
\midrule 
WNh/Slash & 17 & 13 (76.5\%) & 13 (76.5\%) & 0 (0.0\%) & 4 (23.5\%) \\
O dwarfs & 200 & 106 (53.0\%) & 104 (52.0\%) & 5 (4.5\%) & 89 (44.5\%) \\
O giants & 110 & 50 (45.5\%) & 44 (40.0\%) & 13 (20.6\%) & 47 (42.7\%) \\
O no LC & 38 & 17 (44.7\%) & 14 (37.8\%) & 4 (19.0\%) & 17 (44.7\%) \\
B dwarfs & 326 & 189 (58.0\%) & 31 (9.5\%) & 9 (4.6\%) & 128 (39.3\%) \\
B giants & 112 & 69 (61.6\%) & 35 (31.3\%) & 4 (5.5\%) & 39 (34.8\%) \\
WR & 6 & 4 (66.7\%) & 4 (66.7\%) & 0 (0.0\%) & 2 (33.3\%) \\
Later types & 92 & 4 (4.4\%) & 2 (2.2\%) & 0 (0.0\%) & 88 (95.7\%) \\
\midrule
Total & 901$^\star$ & 452 (50.2\%) & 247 (27.4\%) & 35 (7.2\%) & 414 (45.9\%) \\
\bottomrule
\end{tabular}\\
\small
$^\star$ 934 including the remaining 31 \argus IFU (integral-field-unit) targets and VFTS\,338 and~416
\end{table}

\section{Inferring age and mass distributions}\label{sec:inferring-age-mass-distributions}

\begin{figure}[t]
\centering
\includegraphics[width=1.0\textwidth]{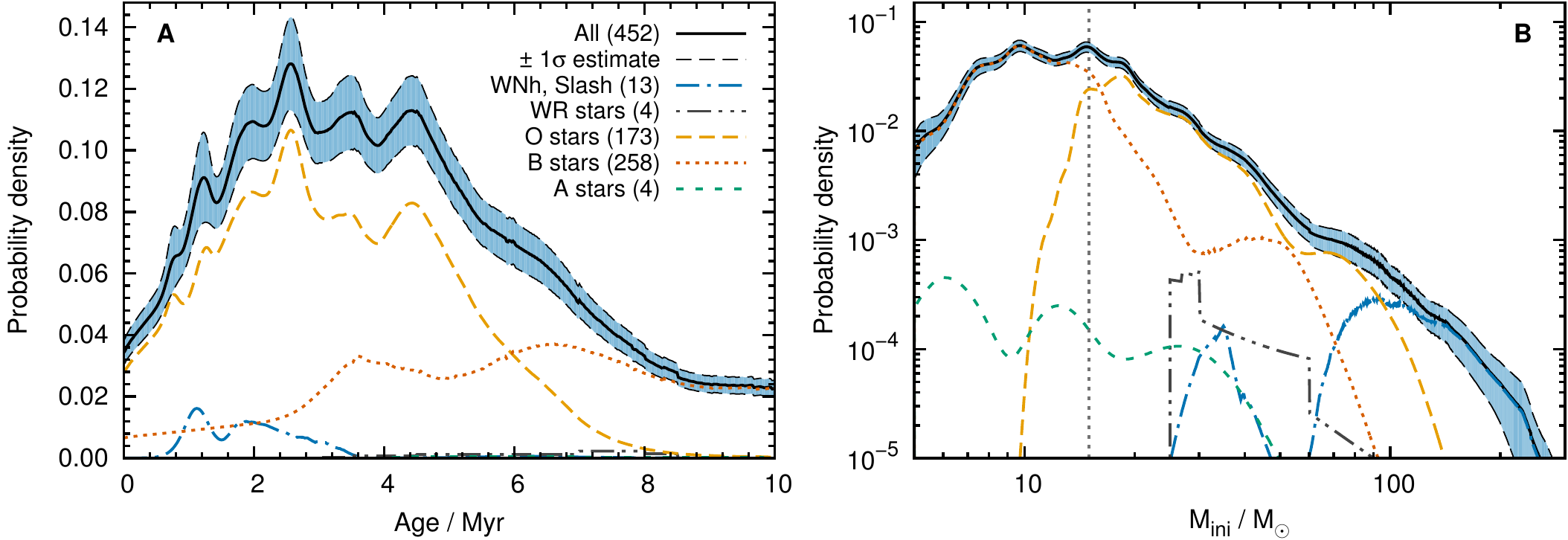}\\
\vspace{4mm}
\includegraphics[width=1.0\textwidth]{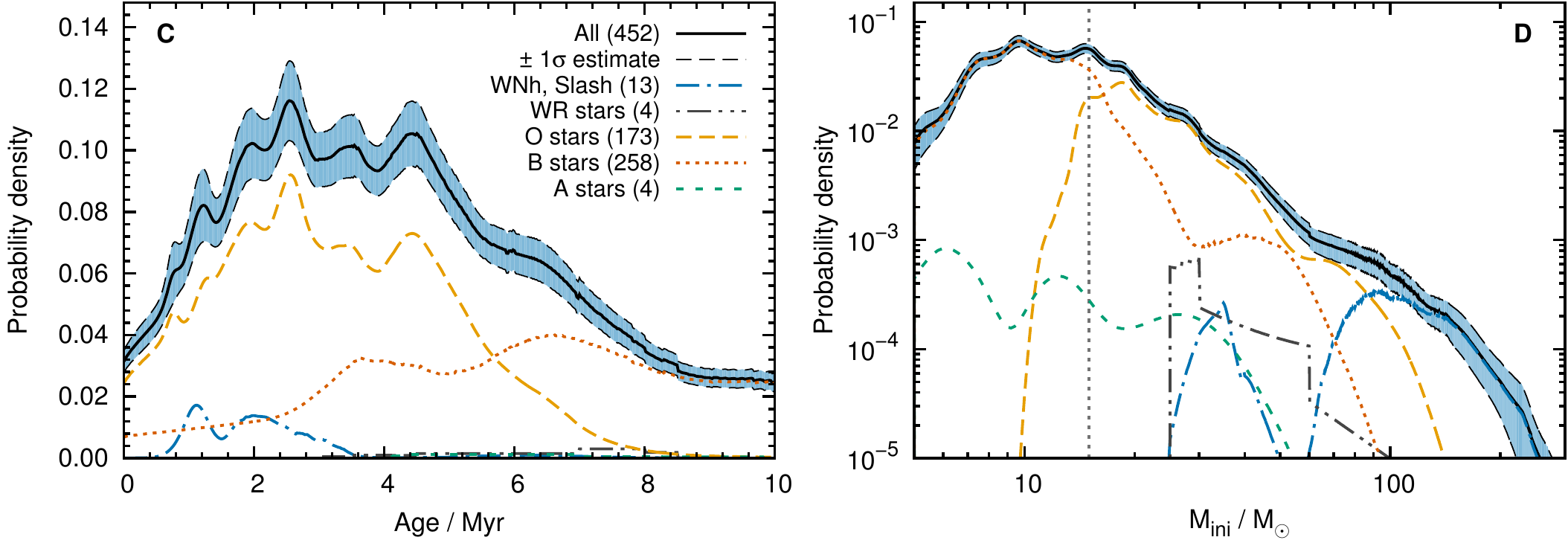}
\caption{\textbf{Age and mass distributions of the 452 VFTS stars in our full sample.} In panels (A) and (B), we apply the completeness corrections described in Sect.~\ref{sec:inferring-age-mass-distributions} while we do not apply them in panels (C) and (D). The contributions of stars of different spectral types are shown and the vertical, grey-dotted lines at $15\,\msun$ in panels (B) and (D) indicate where the mass distributions level off because of the magnitude limit of the VFTS.}
\label{fig:age-mass-distr-all-stars}
\end{figure}

For each star in our full sample, we now have posterior probability distributions of age and initial mass. Summing up the individual distributions gives equivalent distributions for samples of stars. By constructing the distribution of initial masses of our full sample of 452 stars, we confirm that we have good completeness down to masses of $15\,\msun$ because the mass function only begins to level off at lower masses (Fig.~\ref{fig:age-mass-distr-all-stars}). To avoid biases because of an incomplete sample, we thus only work with the 247 stars that are more massive than $15\,\msun$ to derive the SFH and IMF of \tdor. 

To quantify the robustness of the derived distributions and the significance of individual features with respect to the sample size and selection, we estimate $1\sigma$ uncertainties from a bootstrapping technique. We randomly draw, with replacement, 10,000 realisations of 247 stars from our final sample and compute the age and mass distributions for each realisation. The given $1\sigma$ uncertainties are then the standard deviations of the probability distributions of the 10,000 realisations.

As described in Sect.~\ref{sec:sample-selection} and evident from Table~\ref{tab:sample-summary-stats}, our final (sub-)sample of VFTS stars suffers from selection effects such that we have to apply the following four corrections:
\begin{itemize}
\item The completeness in our final sample of stars varies with spectral type and luminosity class because of different detected binary fractions, visual multiple fractions, nebular contamination and contamination from nearby sources (Table~\ref{tab:sample-summary-stats}). We consider the spectral classifications WR, WNh/slash, O dwarf, O giant, B dwarf and B giant and scale their contributions to the age and mass distributions according to their respective completeness within the VFTS. We do not correct for the completeness of later-type stars because the two later-type stars in our final sample are not representative of the full sample of later-type stars in \tdor and most later-type stars in the VFTS are actually less massive than $15\,\msun$.
\item Because of the \flames fibre allocation process, regions of higher stellar densities (close to the R136 cluster core) are less complete than lower surface density regions, requiring a spatial incompleteness correction. To that end, we compute the spatial completeness of massive stars in the VFTS as a function of radial distance to the R136 cluster core using the stellar census of \tdor as a reference distribution \cite{2013A&A...558A.134D}. The spatial completeness is then used to scale the contribution of each star to the age and mass distributions.
\item The \argus data set within the VFTS contains 37 stars. Only the emission-line objects in this subset of stars (four WNh and slash stars, and two O giants) have been analysed so far. We correct for this bias by accordingly increasing the contribution of non-emission-line objects in our age and mass distributions.
\item The sample of \cite{2014A&A...570A..38B} contains the $190\,\msun$ O-supergiant Mk\,42 which is not part of the VFTS sample and is therefore not considered in this work.
\end{itemize}
The four corrections applied together hardly change the shape of the age and mass distributions (Fig.~\ref{fig:age-mass-distr-all-stars}), and we therefore regard our results to not be affected by the selection process.

\begin{figure}[t]
\centering
\includegraphics[width=0.49\textwidth]{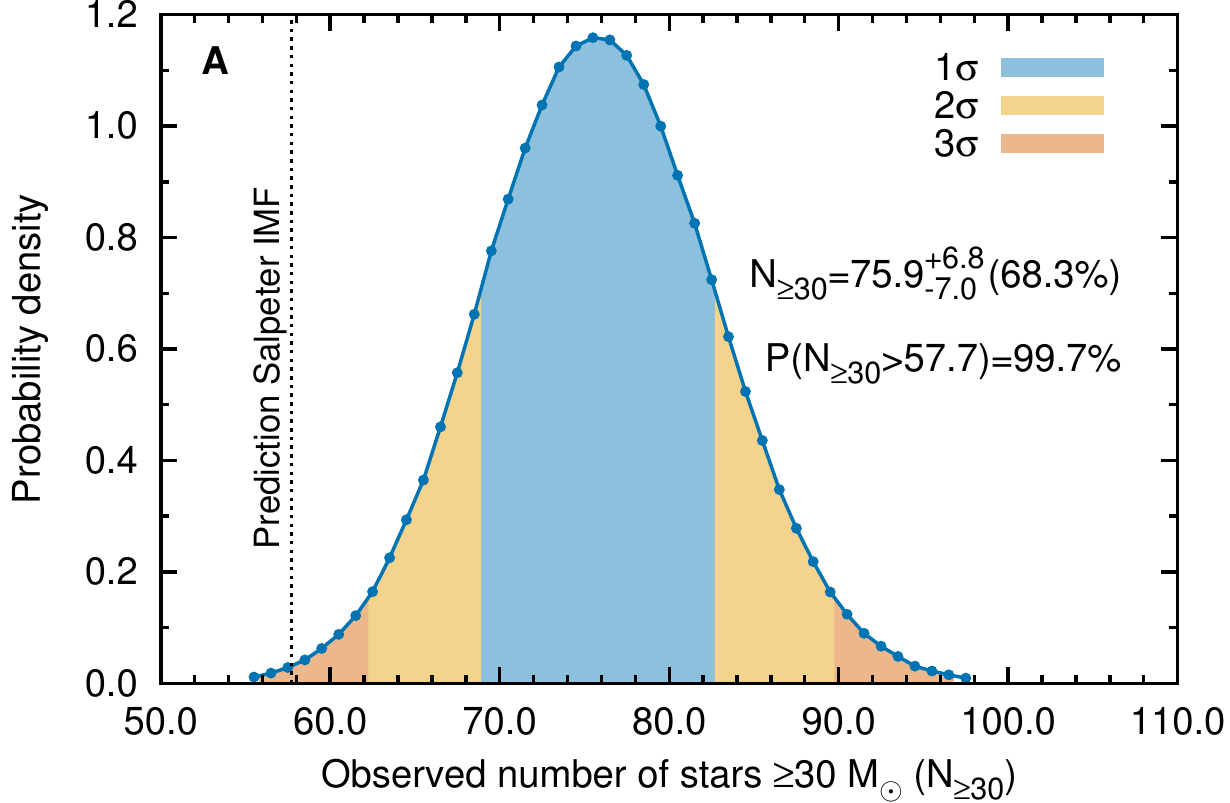}
\includegraphics[width=0.49\textwidth]{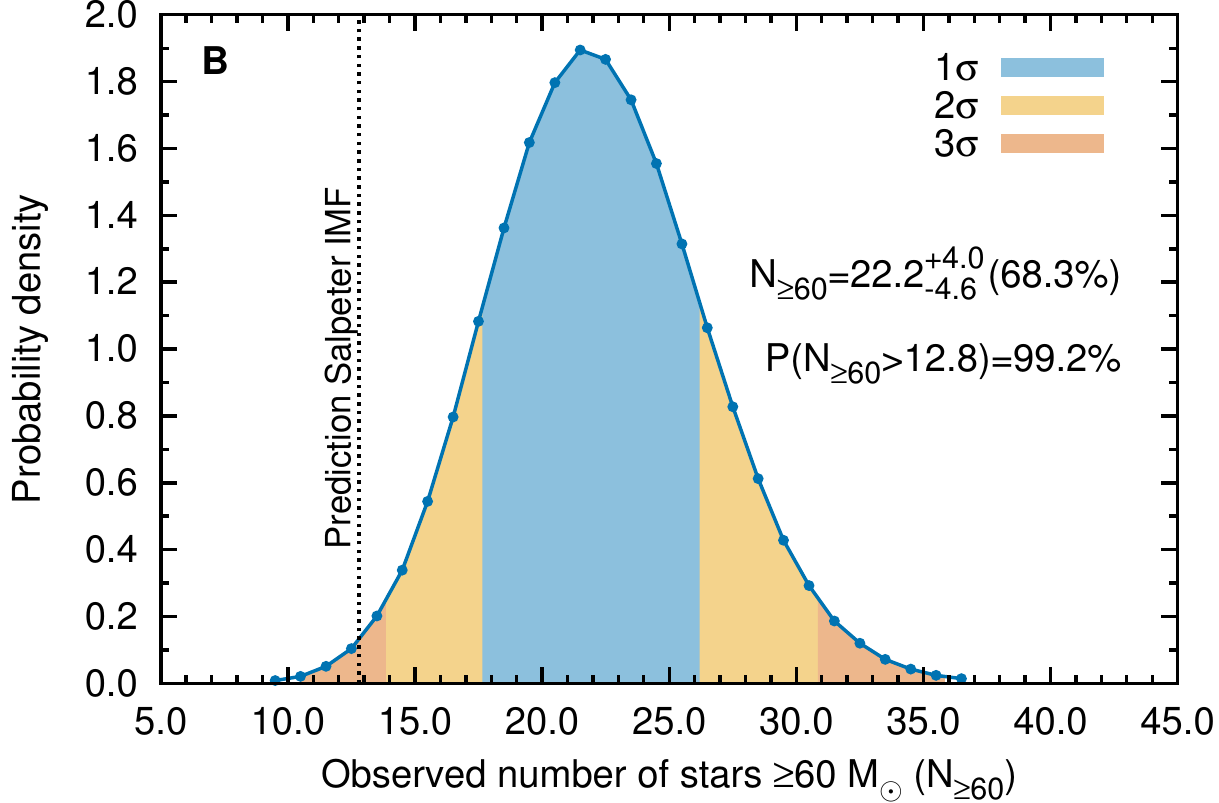}
\caption{\textbf{Probability distribution of the number of stars more massive than (A) $30\,\msun$ and (B) $60\,\msun$.} The predictions of the number of stars more massive than $30\,\msun$ and $60\,\msun$ assuming a Salpeter IMF are indicated by the vertical dashed lines (Sect.~\ref{sec:imf-sfh}). The probabilities that the observed number of stars are larger than the predictions of a Salpeter IMF are $P(N_{\geq 30} > 57.7)=99.7\%$ and $P(N_{\geq 60} > 12.8)=99.3\%$ for masses of $30\,\msun$ and $60\,\msun$, respectively.}
\label{fig:number-of-stars-above-m}
\end{figure}

From the distribution of initial masses of our sample stars and our bootstrapping method, we compute the probability distributions of the number of stars more massive than $30\,\msun$ and $60\,\msun$ (Fig.~\ref{fig:number-of-stars-above-m}). We find $75.9^{+6.8}_{-7.0}$ and $22.2^{+4.0}_{-4.6}$ stars above $30\,\msun$ and $60\,\msun$, respectively. These numbers will be further discussed in Sect.~\ref{sec:imf-sfh}.

\section{Star-formation history and stellar initial mass function}\label{sec:imf-sfh}

Our inferred age and mass distributions are neither star-formation histories nor initial mass functions because we have so far only determined the distributions of ages and initial masses of stars that are still present. In order to derive the SFH and IMF, we have to correct for those stars that already ended nuclear burning. Let $\xi(M)$ be the IMF with $M$ being the initial mass and $S(t)$ the SFH with $t$ the time. Let us also make the usual assumption that the IMF does neither depend on location in \tdor nor age. The probability density functions of ages, $\kappa(t)$, and masses, $\zeta(M)$, of stars observed today are then given by
\begin{equation}
\kappa(t)=\frac{\mathrm{d}p}{\mathrm{d}t}\propto\int_{M_{\mathrm{min}}}^{M_{\mathrm{max}}}\,\xi(M)S(t)\Lambda(t,M)\,\mathrm{d}M
\label{eq:observed-adf}
\end{equation}
and
\begin{equation}
\zeta(M)=\frac{\mathrm{d}p}{\mathrm{d}M}\propto\int_{0}^{T_\mathrm{max}}\,\xi(M)S(t)\Lambda(t,M)\,\mathrm{d}t.
\label{eq:observed-imf}
\end{equation}
The proportionality constants of both $\kappa(t)$ and $\zeta(M)$ follow from normalisation, $\int\,\kappa(t)\,\mathrm{d}t=1$ and $\int\,\zeta(M)\,\mathrm{d}M=1$, respectively. The function $\Lambda(t,M)$ is defined as
\begin{equation}
\Lambda(t,M)=H[\tau(M)-t]=
\begin{cases}
0, & \text{for  } \tau(M)-t<0\\
1, & \text{for  } \tau(M)-t\geq 0
\end{cases}
\label{eq:lambda}
\end{equation}
where $H$ is the Heavyside step-function and $\tau(M)$ the nuclear-burning lifetime of a star. The function $\Lambda(t,M)$ thus describes whether a star of mass $M$ born a time $t$ ago is present today. Here we use the lifetimes of non-rotating, single-star models \cite{2011A&A...530A.115B,2015A&A...573A..71K}. The lifetimes of the rotating models are essentially the same (within a few percent) unless stars rotate initially so rapidly that they evolve chemically homogeneously. The minimum and maximum initial masses of stars to be considered are $M_\mathrm{min}$ and $M_\mathrm{max}$, and the lifetime of stars of the minimum mass sets the maximum age $T_\mathrm{max}$ for which we can reconstruct the SFH. 

Equations~(\ref{eq:observed-adf}) and~(\ref{eq:observed-imf}) further show that the underlying IMF, $\xi(M)$, can only be determined if the SFH is known and vice versa (see also \cite{2006ApJ...636..149E}). For example, with a constant SFH and a power-law IMF, $\xi(M){\propto}M^{-\gamma}$, the mass distribution of stars observed today is $\zeta(M){\propto}M^{1-\gamma-x}$ where $x$ is the exponent of the mass-luminosity relation ($L(M){\propto}M^x$ such that $\tau(M){\propto}M^{1-x}$). The mass-luminosity exponent $x$ approaches $1$ for very massive stars ($M>100\,\msun$) and $\approx 4$ for lower mass stars ($M=1\text{--}2\,\msun$) \cite{2009pfer.book.....M}, showing that the exponent of the mass distribution of stars observed today can be very different from that of the underlying IMF slope $\gamma$. 

Given that we can determine both the age and mass distribution of stars observed today, we can infer the underlying IMF and SFH in \tdor. To find the SFH and IMF, we assume that the IMF has the form of a power-law with slope $\gamma$, $\xi(M)\propto M^{-\gamma}$, that is truncated at $200\,\msun$ (see Sect.~\ref{sec:upper-mass-limit} for a discussion of this upper mass limit). Given an IMF slope, we can compute the SFH for this IMF from Eq.~(\ref{eq:observed-adf}),
\begin{equation}
S(t)=\frac{\kappa(t)}{N\int\,\xi(M)\Lambda(t,M)\,\mathrm{d}M}\label{eq:sfh},
\end{equation}
where $N$ is a normalisation constant (see above). Using this SFH $S(t)$ and the assumed IMF, we compute, from Eq.~(\ref{eq:observed-imf}), the predicted distribution of masses as observed today.

As a first step, we consider a Salpeter IMF with slope $\gamma=2.35$ and its corresponding SFH. Above $30\,\msun$, the simulated mass function appears steeper than that observed and the differences increase with mass (Fig.~\ref{fig:sfh-imf}). A Salpeter IMF predicts $57.7$ ($12.8$) stars above $30\,\msun$ ($60\,\msun$) and therefore underpredicts the number of massive stars by $18.2^{+6.8}_{-7.0}$ ($9.4^{+4.0}_{-4.6}$). Integrating the probability distributions of the number of massive stars in our sample, we find that a Salpeter IMF cannot explain the number of stars above $30\,\msun$ ($60\,\msun$) with 99.7\% (99.2\%) confidence (Fig.~\ref{fig:number-of-stars-above-m}). This allows us to reject the null hypothesis of an IMF slope of $\gamma = 2.35$ for initial masses $\geq 30\,\msun$ at a significance better than 1\%.

We repeat the computations of the SFH and simulated mass functions over a range of adopted IMF slopes, from $\gamma = 1.00$ to $3.50$ in steps of $0.05$. By doing so, we construct a grid of self-consistently derived SFHs and observable mass functions that are normalized to the currently observed population of massive stars ($\geq 15\,\msun$) in \tdor. The simulated distribution of initial masses are then compared to that observed by computing the following two quantities: (i) the number of stars more massive than a mass threshold of $30$ and $60\,\msun$ (Fig.~\ref{fig:number-stars-vs-gamma}), and (ii) the $\chi^2$ between the observed and simulated distributions over the full mass range of $15\,\text{--}200\,\msun$ of our sample stars, using the bootstrapped $1\sigma$ estimates as uncertainties. These procedures then allow us to find the best match between the observed and simulated quantities. For both diagnostics, we compute a probability distribution of the IMF slopes (Figs.~\ref{fig:number-stars-vs-gamma} and~\ref{fig:gamma}). Based on the number of stars more massive than $30\,\msun$ and $60\,\msun$, we find an IMF slope of $\gamma=1.84^{+0.18}_{-0.18}$ and $\gamma=1.84^{+0.22}_{-0.17}$, respectively. Fitting the observed distribution of initial masses over the mass range $15\text{--}200\,\msun$, yields $\gamma = 1.90^{+0.37}_{-0.26}$ (Fig.~\ref{fig:gamma}). Both optimization methods are thus in excellent agreement. We adopt $\gamma = 1.90^{+0.37}_{-0.26}$ as our overall best-fitting IMF slope because it is derived by considering the whole range of masses of our sample stars. This IMF slope then also fixes the best-fitting SFH shown in Fig.~\ref{fig:sfh-imf}.

For each IMF and corresponding SFH, we can compute the relative number of stars that already ended their lives, $N_x$, from,
\begin{equation}
\frac{N_{x}}{N_{\geq M_{\mathrm{min}}}^{\mathrm{today}}}=\frac{\int_{M_{\mathrm{min,x}}}^{M_{\mathrm{max,x}}}\,\int_{0}^{T}\,\xi(M)S(t)\Omega(t,M)\,\mathrm{d}t\,\mathrm{d}M}{\int_{M_{\mathrm{min}}}^{M_{\mathrm{max}}}\,\int_{0}^{T}\,\xi(M)S(t)\Lambda(t,M)\,\mathrm{d}t\,\mathrm{d}M},
\label{eq:fraction-expired-stars}
\end{equation}
where $N_{\geq M_{\mathrm{min}}}^{\mathrm{today}}=247$ is the number of stars initially more massive than $M_\mathrm{min}=15\,\msun$ as observed today in our sample and $\Omega(t,M)$ is a step function stating whether a star of initial mass $M$ born a time $t$ ago ended its life by today (it is the opposite of $\Lambda(t,M)$). By adjusting $M_{\mathrm{min,x}}$ and $M_{\mathrm{max,x}}$, we can also compute the number of stars in the mass interval $M_{\mathrm{min,x}}\text{--}M_{\mathrm{max,x}}$ that ended their lives to \eg obtain the number of stars that exploded in a supernova, formed a neutron star or formed a black hole. For \tdor, we find that 140 stars more massive than $15\,\msun$ ended their lives within the last $12\,\myr$. Of these 140 stars, about 50 exploded in a supernova and about 130 left a black hole behind (assuming that stars with initial masses up to $40\,\msun$ explode in a supernova and stars above $25\,\msun$ leave a black-hole remnant; see Sect.~\ref{sec:feedback}). We can only reconstruct the IMF for stars initially more massive than $15\,\msun$, limiting the SFH to about $t \lesssim 12\,\myr$ and our ability to infer the number of stars that ended their lives to masses $\geq 15\,\msun$.

We assume a single power-law IMF model. The observed distribution of initial masses in Fig.~\ref{fig:sfh-imf}B shows the largest mismatch with a Salpeter IMF slope at the high-mass end ($\gtrsim 30\,\msun$) and it may thus be conceivable that the true IMF is better approximated by a two-part power-law model with a Salpeter slope below about $30\,\msun$ and a flatter slope above. With the current data it is difficult to discriminate between these possibilities. The IMF around R136 has been probed by other authors and is found to be consistent with a Salpeter IMF slope below $\approx 20\,\msun$ \cite{2009ApJ...707.1347A,2015ApJ...811...76C}. Cignoni et al. further conclude \cite{2015ApJ...811...76C}: ``At high masses, our synthetic [color-magnitude diagrams] tend to underestimate the star counts in the densest regions. This may suggest a flattening of the IMF above $10\,\msun$.''

It is noteworthy that our inferred IMF slope of massive stars in \tdor is close to the asymptotic limit of $\gamma \rightarrow 2.00$ expected for stars that have formed via gravitationally focussed mass accretion with mass accretion rates proportional to mass squared, $\dot{M}\propto M^2$, \ie Bondi--Hoyle--Littleton like accretion \cite{1982NYASA.395..226Z,2015MNRAS.452..566B}. This limit may only be reached if stars grow well beyond their initial seed mass \cite{1982NYASA.395..226Z} which could be the case for the massive stars in our sample. If this mode of star formation is responsible for the overabundance of massive stars found in this work in \tdor, it would be a universal feature of star formation that the IMF slope approaches a value of $\gamma=2.00$ at the high mass end. This limit would not be reached at low mass where stars do not accrete a substantial fraction of their seed mass such that the IMF slope might transition from a Salpeter-like slope of $\gamma=2.35$ at lower masses to the asymptotic limit of $\gamma=2.00$ at higher masses.

\clearpage
\section*{Supplementary Text}

We discuss potential systematics that may influence the inference and interpretation of the SFH and IMF (Sect.~\ref{sec:discussion}). We then put constraints on the maximum birth mass of stars (Sect.~\ref{sec:upper-mass-limit}) and estimate the increase in various feedback properties from stellar populations with an IMF slope flatter than Salpeter and a variable upper-mass limit (Sect.~\ref{sec:feedback}).

\section{Discussion of potential biases in the inferred SFH and IMF}\label{sec:discussion}

In this section, we discuss several systematics that may influence the inference and/or interpretation of our SFH and IMF of massive stars in \tdor. In particular, we consider the following aspects:
\begin{itemize}
\item We remove all known binaries from our sample, but there will be some left that can affect our results. The influence of such unrecognised binaries on our sample is examined in Sect.~\ref{sec:unresolved-binaries}.
\item About $15\text{--}40\%$ of our sample stars are expected to be products of binary mass transfer \cite{2014ApJ...782....7D}, \ie they have accreted mass in a past mass exchange episode and/or merged with a former binary companion. We discuss the influence of binary mass transfer on the inferred IMF in Sect.~\ref{sec:binary-mt}.
\item The only known region in \tdor that has not been observed within the VFTS and that contains a significant fraction of stars more massive than $15\,\msun$ is the R136 star cluster. We therefore discuss whether runaways ejected from R136 and entering our sample could have affected our interpretation of the observed mass distribution of massive stars in \tdor (Sect.~\ref{sec:r136-runaways}). Furthermore, we provide an estimate of the IMF of stars in the core of R136 (Sect.~\ref{sec:r136-imf}) to investigate how the omission of R136 might influence our results.
\item The atmospheric and hence fundamental stellar parameters of the B dwarfs and some B giants are biased because we applied atmosphere models with an offset in the metallicity compared to that of stars in the LMC (Sect.~\ref{sec:pd-ob-stars}). This aspect is further discussed in Sect.~\ref{sec:discussion-b-stars}.
\item Massive stars are not yet fully understood and the stellar evolution models likely do not incorporate all the relevant physics that could influence the inference of masses and ages, and hence the IMF slope. We discuss some of these aspects in Sect.~\ref{sec:discussion-models}.
\end{itemize}

\subsection{Unresolved binaries}\label{sec:unresolved-binaries}

Unresolved binaries and other multiple stellar systems can bias the inference of the IMF \cite{1991A&A...250..324S,1993MNRAS.262..545K,2008ApJ...677.1278M,2009MNRAS.393..663W,2015ApJ...805...20S}. The larger luminosities associated with binary stars can result in overestimated stellar masses and hence an apparent flattening of the inferred IMF. A key point is that the mass-luminosity (ML) relation of stars, $L \propto M^x$, depends on mass ($L$ is the luminosity of stars, $M$ the stellar mass and $x$ the ML exponent). The ML exponent $x$ is smaller for larger masses such that the mass inferred of an equal-mass binary from the combined luminosity of both stars, $M_\mathrm{obs}=2^{1/x} M$, is larger for smaller $x$, \ie at higher masses. At higher masses, the inferred IMF is thus stronger affected by this, resulting in the aforementioned flattening. If the ML relation was not mass dependent, the inferred IMF slope would remain unchanged.

In the VFTS, we have excluded known binaries such that the above bias is minimised. The fraction of unrecognised binaries $f_\mathrm{B}^\mathrm{unrec}$ in a sample that has an intrinsic binary fraction of $f_\mathrm{B}^\mathrm{int}$ and a binary detection fraction of $f_\mathrm{B}^\mathrm{det}$ is
\begin{equation}
f_\mathrm{B}^\mathrm{unrec} = \frac{f_\mathrm{B}^\mathrm{int}-f_\mathrm{B}^\mathrm{det}}{1-f_\mathrm{B}^\mathrm{det}}.
\label{eq:unrec-binary-fraction}
\end{equation}
In the VFTS, the intrinsic binary fraction of O stars is found to be $f_\mathrm{B}^\mathrm{int}=0.51\pm0.04$ and the binary detection fraction is $f_\mathrm{B}^\mathrm{det}=0.35\pm0.03$ \cite{2013A&A...550A.107S}. This means that about 25\% of our VFTS sample stars are unrecognised binaries.

\begin{figure}[t]
\centering
\includegraphics[width=1.0\textwidth]{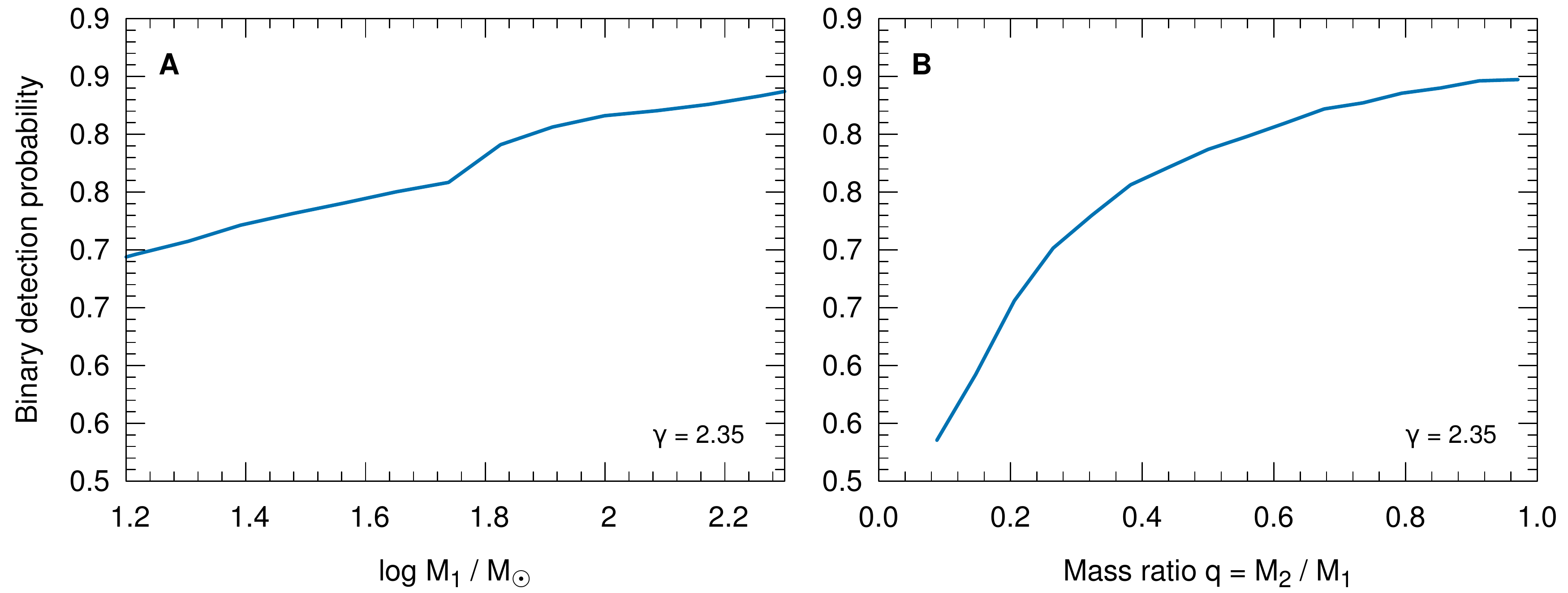}
\caption{\textbf{Binary detection probability of the VFTS as a function of (A) the mass of the primary star, $M_1$, and (B) the mass ratio, $q=M_2/M_1$, where $M_2$ is the mass of the secondary star ($M_2<M_1$).} The kink in the detection probability as a function of primary mass at $80\,\msun$ ($\log M_1/\msun \approx 1.8$) is an artefact as we transition from binary detection rates of O stars to emission line objects abruptly at this mass. Note that these detection probabilities are characteristic of the VFTS \cite{2013A&A...550A.107S} and not directly applicable to other spectroscopic surveys.}
\label{fig:binary-detection-probability}
\end{figure}

In a spectroscopic survey such as the VFTS, binaries are identified by their radial-velocity (RV) variations. Such RV variations are largest in binaries with the most massive primary stars $M_1$, the largest mass ratios $q=M_2/M_1$ (where $M_2$ is the mass of the secondary, $M_2<M_1$), the shortest orbital periods and smallest eccentricities (see \eg fig.~8 in Ref.~\citenum{2013A&A...550A.107S}). In Fig.~\ref{fig:binary-detection-probability}, we have computed the binary detection probabilities of stars in Refs.~\citenum{2013A&A...550A.107S} and~\citenum{2014A&A...570A..38B}, which incorporate the variable accuracy of the RV measurements as a function of the signal-to-noise ratio, rotation rate, spectral shape and time sampling of the VFTS data. When removing identified binaries from our sample, we therefore preferentially remove binaries at high masses and at large mass ratios.

To study the influence of unresolved binaries quantitatively, we conduct two experiments. First, we sample a population of single and binary stars for a fixed binary fraction of 25\% as expected for our sample. Second, we sample a stellar population with a binary fraction of 51\%, apply the VFTS binary detection probabilities (Fig.~\ref{fig:binary-detection-probability}) and remove the identified binaries from the sample. The two stellar populations therefore have the same binary fractions but the binaries are distributed differently in terms of primary and secondary mass. This will allow us to disentangle biases induced only by the ML relation from those induced by the VFTS binary detection probabilities. We assume that single star masses and the masses of primary stars in binaries are drawn from the same power-law mass function with slope $\gamma$ and that the mass ratios of binaries are sampled from a distribution function of the form $f_q \propto q^\kappa$ with $\kappa=-1.0$ as found for O stars in the VFTS \cite{2013A&A...550A.107S}. Also orbital periods, eccentricities and inclinations of the binary orbits are sampled as found in the VFTS. In our experiments, we only study zero-age MS stellar populations and compute the IMF of all single and binary stars from their known true masses (called ``true IMF'' from here on) and the ``observed IMF'' by converting the total luminosity of binary stars into an ``observed'' mass by inverting the ML relation of the stellar models of Ref.~\citenum{2011A&A...530A.115B}. These two IMFs are then fitted by power-law functions with a least-squares algorithm to infer differences in the inferred IMF slopes. We sample 5000 stars and repeat the sampling experiment 10,000 times to obtain variations in the IMFs and inferred IMF slopes. We are not sampling the same number of stars as in the VFTS because we are not interested in effects because of stochastic sampling but only in effects because of unresolved binaries.

\begin{figure}[t]
\centering
\includegraphics[width=0.75\textwidth]{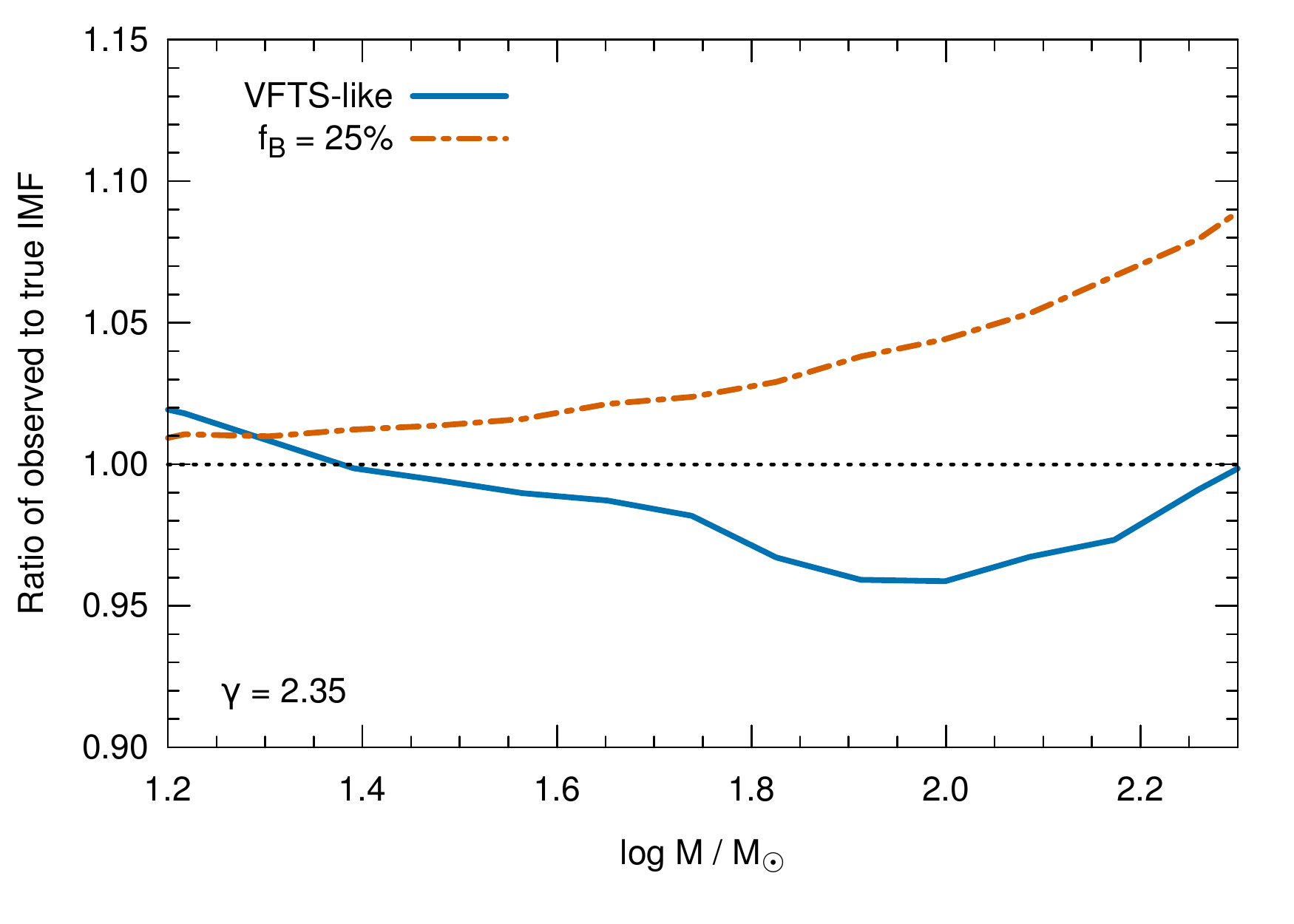}
\caption{\textbf{Ratio of the observed to the true IMF.} The solid blue curve shows the ratio when taking the VFTS binary detection probability properly into account whereas the red dot-dashed curve shows the ratio of observed to true IMF for a stellar population with a binary fraction of 25\% without using the VFTS binary detection probabilities. The true IMF of stars has a slope of $\gamma=2.35$ and the black dotted line shows the one-to-one ratio of observed and true IMF.}
\label{fig:imf-ratio}
\end{figure}

In Fig.~\ref{fig:imf-ratio}, we show the ratios of the observed to the true IMF of our two experiments. In the first experiment, the IMF flattens, \ie the ratio increases with mass, because of the bias induced by the unresolved binaries and the ML relation as discussed above. The inferred IMF slope is flatter than the true IMF by $0.024\pm0.018$. This can be viewed as an upper limit because we have only sampled zero-age MS populations and taking the full star formation history into account would reduce this bias (see below). When also taking the VFTS binary detection probability properly into account, we are more efficient in removing binaries at high mass than at low mass, steepening the observed IMF (Fig.~\ref{fig:imf-ratio}). Furthermore, the remaining unrecognised binaries preferentially have low mass ratios where the bias because of the ML relation is small. Overall, the observed IMF slope steepens by $0.034\pm0.039$ compared to the true IMF. In our work, we apply incompleteness corrections as a function of spectral type to correct for the different completeness levels of our sub samples. This somewhat reduces the effect of removing more binaries at high than at low masses (Fig.~\ref{fig:binary-detection-probability}A).

It is worthwhile to realise the following limitations of the experiments discussed here. 
\begin{itemize}
\item We have assumed that the single and primary stars in binaries follow the same IMF. To our knowledge, there is yet no conclusive evidence that supports or contradicts this assumption. If the single and primary stars would follow different IMFs, the results presented in this work would still remain valid for single stars.
\item Another simplification is to only sample zero-age MS populations. This eases the experiments because we do not need to worry about wind mass loss, binary mass exchange and the star formation history. Still, the general behaviour of the inferred IMF because of potential biases from unrecognised binaries becomes evident from our experiments. If we were to also take the star formation history of \tdor into account, the bias from unresolved binaries would become weaker. This is because the IMF is most strongly biased at the high mass end, \ie by the youngest stars, and the star formation history of our sample of VFTS stars puts more weight on mass ranges of the IMF where the bias from unresolved binaries is weaker than at the high mass end (Fig.~\ref{fig:imf-ratio}).
\item In a spectroscopic survey, binary stars will not only bias luminosities but also other atmospheric parameters derived from composite spectra. Mimicking these effects is much more difficult but the main bias is via the luminosity that constrains inferred stellar masses strongest. The fact that unidentified binaries in a VFTS-like spectroscopic survey are predominantly composed of binaries with low-mass companions, lessens their impact on the inferred atmospheric parameters of the primary stars. Low mass companions indeed hardly contribute to the total flux, which is one of the reasons why they are harder to detect.
\end{itemize}

In conclusion, we find that unrecognised binaries hardly bias the inference of the IMF in our case and, if at all, the bias seems to be such that the true IMF slope might be even flatter than what we infer. This is because the VFTS is quite efficient in identifying binaries. The remaining binaries are too few to significantly affect our conclusions.

\subsection{Binary mass transfer}\label{sec:binary-mt}

Past episodes of binary mass transfer may affect our results in two ways. First, binary mass transfer produces a surplus of massive stars. In coeval stellar populations where the mass function is truncated at the turn-off mass (the mass of the most massive star that has not yet ended nuclear burning), binary mass transfer adds a tail of binary products (blue stragglers) that extends the mass function by up to a factor of 2 in mass. This binary tail is less populated than the original IMF \cite{2015ApJ...805...20S}. Second, mass accretion rejuvenates stars such that they look younger than they really are \cite{1983Ap&SS..96...37H,1995A&A...297..483B,2016MNRAS.457.2355S}. 

Generally, if the initial masses of some stars in our sample are overestimated, \eg, because of a past binary mass-transfer episode, the real IMF is steeper than that inferred (it is flatter if some masses are underestimated). Similarly, if the ages of some of our stars are underestimated, \eg, because of rejuvenation, the real IMF slope is flatter than that inferred (it is steeper if ages are overestimated). The latter is true because older stars are, on average, less massive because less massive stars have longer lifetimes (and vice versa). Consequently, binary mass transfer produces a surplus of massive stars that biases the inferred IMF slope to flatter values whereas the associated rejuvenation leads to underestimated ages, biasing the inferred IMF slope to steeper values.

In order to quantify how binary mass transfer affects the inference of the IMF slope, we add a tail that extends the original IMF by a factor of 2 in mass at a reduced level of 20\%. These numbers are based on detailed population synthesis models \cite{2015ApJ...805...20S}. Rejuvenation is modelled by assuming that our sample contains 30\% of rejuvenated binary products \cite{2014ApJ...782....7D} which appear 30\% younger than they really are \cite{2016MNRAS.457.2355S}. 

Only adding the above mentioned tail to the IMF model steepens the inferred IMF slope by $0.10$, from $\gamma=1.90$ to $\gamma=2.00$ and only considering rejuvenation by modifying the observed age distribution flattens the inferred IMF slope by the same amount of $0.10$ to $\gamma=1.80$. Considering both effects at the same time, we find a best-fitting IMF slope of $\gamma=1.90^{+0.35}_{-0.25}$. We therefore conclude that, in our case, binary mass transfer and the associated rejuvenation cancel out each other's effects on the inference of the IMF slope and are thus not responsible for the apparently shallow IMF slope in \tdor.

\subsection{Runaways from the R136 star cluster}\label{sec:r136-runaways}

With an age of $1\text{--}2\,\myr$ \cite{2016MNRAS.458..624C}, the R136 star cluster is most likely too young to have produced runaway stars by supernova binary disruption but could have produced runaways by cluster dynamical ejection. Such runaway stars are expected to be biased towards high masses \cite{1961BAN....15..265B,2011Sci...334.1380F,2015ApJ...805...92O} which might help to explain the large number of massive stars found in the \tdor field. Runaways that originate from other parts in \tdor and formed by the supernova ejection mechanism do not bias our sample because all other dense regions in \tdor containing massive stars are represented in our sample and have already been accounted for when discussing the potential impact of binary mass transfer on our results (Sect.~\ref{sec:binary-mt}).

We find $18.2^{+6.8}_{-7.0}$ excess stars more massive than $30\,\msun$ compared to a Salpeter IMF. The VFTS is complete to about 73\% (Fig.~\ref{fig:vfts-completeness}) such that we expect about $18.2/0.73\approx 25$ apparently single, excess stars above $30\,\msun$ in the whole of \tdor. In principle, also those VFTS stars excluded from our sample should contain such massive objects, suggesting that there are even more excess stars: our full sample contains about 50\% of all VFTS stars (Table~\ref{tab:sample-summary-stats}) such that we may expect to find up to 50 excess stars in the whole of \tdor. 

In comparison, some N-body simulations of massive star clusters produce about 5 O-type runaways \cite{2011Sci...334.1380F} whereas others predict about 10 O-type runaways for a massive cluster of $5\times10^4\,\msun$ that is comparable to R136 \cite{2015ApJ...805...92O}. More massive clusters could have produced more runaways. Not all of these O-type runaways are single stars and not all of them are more massive than $30\,\msun$. Assuming that 70\% of them are in fact more massive than $30\,\msun$ (corresponding to a mass function slope of $\gamma \approx 1.10$ for the runaways; the fraction is about 40\% and 50\% for a Salpeter IMF slope and $\gamma=1.90$ as we found in \tdor, respectively), we may expect 4--7 runaways more massive than $30\,\msun$ of which some may be binary stars. This is at odds with the 36--71 dynamically-ejected O-type runaways ($\geq 15\,\msun$) needed to explain the large number of 25--50 massive excess stars ($\geq 30\,\msun$) inferred in \tdor. Given these numbers and current N-body models of massive clusters, we consider it unlikely that such runaways from R136 are the cause of the shallow IMF slope found for the VFTS stars in \tdor.

From an observational point of view, there are six known candidate radial-velocity runaways \cite{2014A&A...564A..40W,2015A&A...574A..13E} in our sample that have initial masses $\gtrsim 30\,\msun$ and that, given their ages of $\lesssim 2\,\myr$, could originate from R136: VFTS\,016, 072, 355, 418, 755 and 797. The origin of these six candidate runaways in \tdor is, however, unknown as they have not yet been identified to be runaways by proper-motion studies \cite{2015AJ....150...89P}. The fast projected rotational velocities of VFTS\,072 and 755 (about $185\,\kms$ and $285\,\kms$, respectively; Table~\ref{tab:stellar-parameters}) may suggest that these objects accreted mass from a former binary companion that disrupted the binary by a supernova explosion leading to their ejection rather than being dynamically ejected from R136. Identifying slower runaways, \ie $\lesssim 30\,\kms$ for radial-velocity \cite{2014A&A...564A..40W,2015A&A...574A..13E} and $\lesssim 50\,\kms$ for proper-motion candidates \cite{2015AJ....150...89P}, is difficult. Hence, there remains uncertainty in the true number of runaway candidates from R136 above $30\,\msun$ but the current numbers are consistent with the above mentioned theoretical expectations of cluster dynamics and thus unlikely to explain the large number of massive stars found in \tdor.

It has been suggested that R136 may be in the process of merging with stars in the north-east clump \cite{2012ApJ...754L..37S}. Such a merger may produce a large number of runaways. However, in a merger of two clusters, stars with the lowest binding energies, \ie lowest masses, are preferentially ejected such that the mass function of ejected stars is steeper than their IMF. If R136 is indeed merging with the north-east clump and, if our sample contains stars ejected in this way, our inferred IMF slope would be biased towards steeper values and the real IMF would be flatter than what we find.

\subsection{Mass function of stars in the R136 cluster core}\label{sec:r136-imf}

The young R136 star cluster contains massive stars that are not in our VFTS sample. Thanks to Hubble Space Telescope observations, the core region of the R136 star cluster (the innermost $0.5\,\mathrm{pc}$ around R136a1) has been observed, yielding first estimates of the ages and masses of massive stars therein \cite{2016MNRAS.458..624C}. However, we have not included these stars in our sample because the massive star population in the core of R136 is likely biased in as yet unknown ways. Dense star clusters such as Arches and Quintuplet in the Galactic Centre show evidence of mass segregation that flattens the apparent IMF of stars in their core regions \cite{2010MNRAS.409..628H,2013A&A...556A..26H} and this may be relevant for R136, too. Also, R136 could have produced runaways and/or be in the process of merging with the north-east clump (Sect.~\ref{sec:r136-runaways}), both affecting the mass function of stars in the R136 core. Because of these uncontrollable biases and a different completeness with stellar mass than for stars in the VFTS, we do not include the R136 stars in our analysis.

\begin{figure}[t]
\centering
\includegraphics[width=0.8\textwidth]{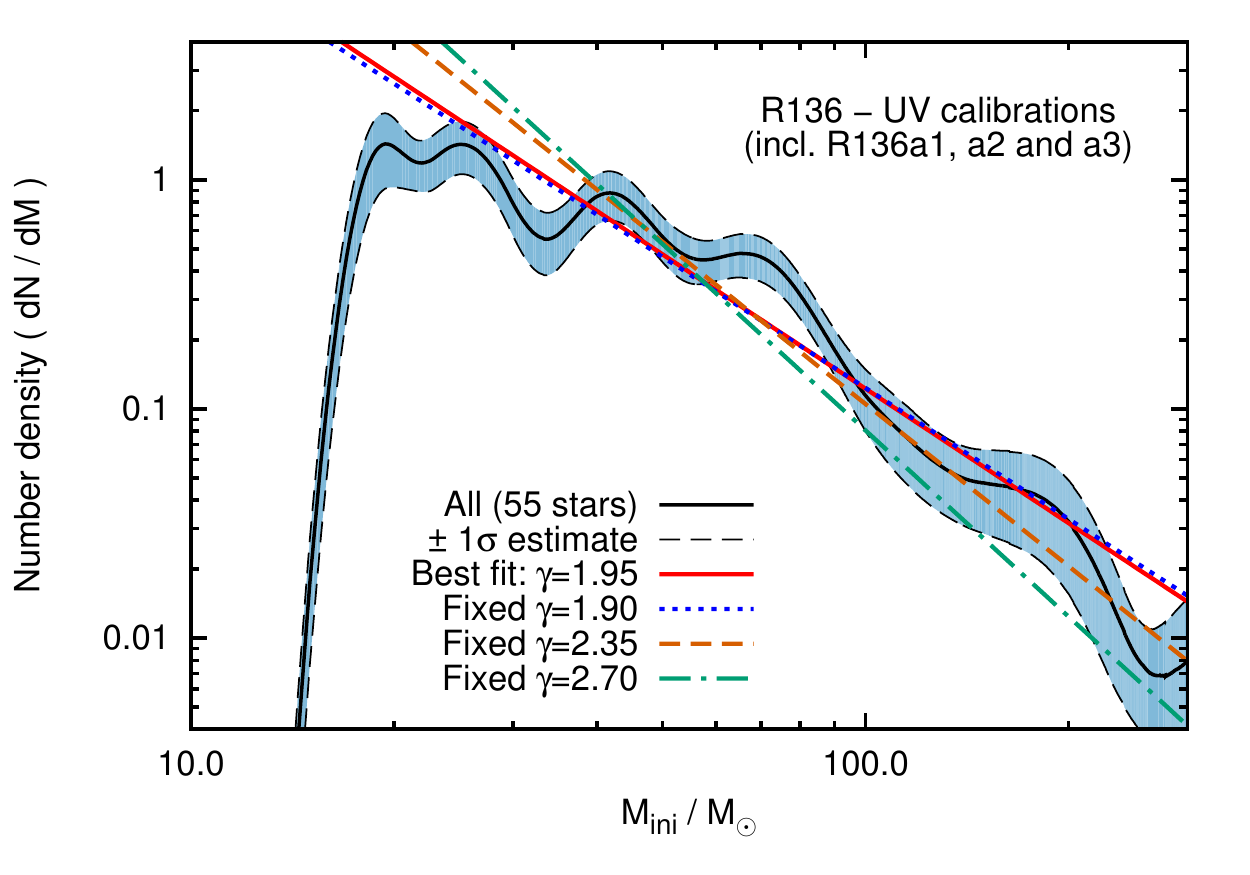}
\caption{\textbf{Distribution of initial masses of 55 stars in the core of the R136 star cluster.} As in Fig.~\ref{fig:sfh-imf}, the shaded region indicates bootstrapped $1\sigma$ estimates. Power-law mass functions are fitted to the distribution over the mass range $30\text{--}200\,\msun$. Because of the large uncertainties and the resulting sensitivity of the fits to the fitted mass range, we also provide IMF models for fixed power-law exponents along with the best-fitting IMF for reference.}
\label{fig:r136-imf}
\end{figure}

Still, the mass function of stars in the core of R136 may hold important information and we can study it separately from our analysis of stars in the surrounding fields of \tdor. Using the inferred stellar parameters of \cite{2016MNRAS.458..624C} from spectral calibrations (luminosity and effective temperature), we compute the distribution of initial masses of stars in the R136 core in the same way as for our stars in \tdor (Fig.~\ref{fig:r136-imf}), \ie assuming the same prior distributions and stellar models. A correction for stars that already ended nuclear burning is not necessary because R136 is so young that all stars are still present. 

The observations of the R136 core \cite{2016MNRAS.458..624C} comprise 55 stars, among them the very massive stars R136a1, a2 and a3 found to exceed initial masses of $150\,\msun$ \cite{2010MNRAS.408..731C}. The sample is unbiased for stars with masses larger than about $30\,\msun$ (fig.~10 in \cite{2016MNRAS.458..624C}) and we thus fit power-law mass functions to the data over the mass range $30\text{--}200\,\msun$. Unfortunately, the mass uncertainties of individual stars are large because of uncertain stellar parameters (the parameters are only estimated from spectral type calibrations), which translates into large uncertainties in the mass distribution. This hampers our ability to infer robust IMF slopes (Fig.~\ref{fig:r136-imf}). We therefore conclude that the IMF of stars in the core of R136 is consistent with the inferred shallow IMF slope of other stars in \tdor but also with a Salpeter IMF slope of $\gamma=2.35$. However, as described above, we expect that the mass function of stars in the R136 core does not necessarily reflect the IMF, complicating its interpretation. Further intricacies may arise when considering the possibility that star formation in very dense regions proceeds differently from that in less dense fields \cite{1992Natur.359..305P,2007MNRAS.381L..40D}.

\subsection{B stars}\label{sec:discussion-b-stars}

Because of the metallicity offset of the atmosphere grid used to analyse some B stars (Sect.~\ref{sec:pd-ob-stars}), their effective temperatures can be too cool by $2000\,\mathrm{K}$ for $T_\mathrm{eff}<25,000\,\mathrm{K}$ and too hot by $1000\,\mathrm{K}$ for $T_\mathrm{eff}>25,000\,\mathrm{K}$. This directly translates into differences in the inferred luminosities (hotter temperatures result in larger luminosities and vice versa) and hence stellar ages and masses (hotter temperatures give younger ages and larger masses, and vice versa). These systematics only apply to the stellar parameters of B dwarfs and some of the B giants as detailed in Sect.~\ref{sec:pd-ob-stars} (see also Table~\ref{tab:sample-summary-stats}).

The B stars in our final sample are found to be slightly more massive than $15\,\msun$ (Table~\ref{tab:stellar-parameters}) and do not affect the derived IMF at masses $\gtrsim 20\,\msun$. The high mass end of the IMF, where we find the strongest deviation with respect to a Salpeter IMF model, is therefore not affected by the bias present in the derived stellar parameters of these B stars. We further quantify this by deriving the SFH and IMF only from stars more massive than $20\,\msun$. This new mass cut inevitably also removes several O-stars from our sample and the sample size shrinks to 145 stars. This corresponds to a loss of about 40\% of our sample and thus influences the significance of the inferred SFH and IMF. Using the $20\,\msun$ mass cut, we find an IMF slope of $\gamma=1.90^{+0.42}_{-0.32}$ compared to $\gamma=1.90^{+0.37}_{-0.26}$ for a mass cut of $15\,\msun$, showing that our main conclusions remain untouched.

\subsection{Massive star models}\label{sec:discussion-models}

The strongest deviation from a Salpeter high-mass IMF is found at large stellar masses ($\geq 30\,\msun$; Fig.~\ref{fig:sfh-imf}). Because of the scarcity of such massive stars, it is difficult to probe and constrain high-mass stellar evolution models with observations. 

The massive star models used in this work \cite{2011A&A...530A.115B,2015A&A...573A..71K} develop inflated envelopes at initial masses of $\gtrsim 50\,\msun$ \cite{2015A&A...580A..20S} such that stars reach cooler temperatures on the main-sequence than models with less inflated envelopes \cite{2013MNRAS.433.1114Y}. The luminosity evolution remains largely unaffected by inflation. Inflation does therefore not greatly affect initial masses inferred from the position of stars in the Hertzsprung--Russell diagram which are mostly controlled by luminosity. For models with less inflated envelopes (which are typically hotter during the main sequence), the difference in effective temperature leads to older estimated ages. Assuming that the ages of all stars more massive than $50\,\msun$ are systematically increased by $0.5\,\myr$, we find that the inferred IMF slope of our stars in \tdor flattens by $0.05$ to $\gamma=1.85^{+0.38}_{-0.25}$.

In the most massive VFTS stars ($\gtrsim 80\text{--}90\,\msun$), the wind mass-loss rates are found to increase in massive stars that develop optically thick winds \cite{2014A&A...570A..38B}, a finding that has also been theoretically predicted \cite{2008A&A...482..945G,2011A&A...531A.132V}. This implies that the wind mass loss and hence the inferred initial masses of these massive stars has been underestimated. Furthermore, an enhanced wind pushes stellar models to hotter temperatures and we would have underestimated their ages as well. Assuming again that stars $\geq 50\,\msun$ are $0.5\,\myr$ older and that stars $\geq 90\,\msun$ have 15\% larger initial masses because of underestimated winds, we find that the inferred IMF slope flattens by $0.1$ to $\gamma=1.80^{+0.42}_{-0.28}$. Enhanced winds also increase the inferred stellar upper-mass limit (Sect.~\ref{sec:upper-mass-limit}).

\section{Stellar upper-mass limit}\label{sec:upper-mass-limit}

Whether or not there exists a limit to the maximum birth mass of stars, and if so, which physics governs it, are open questions. In the past, a stellar upper-mass limit of $150\,\msun$ has been suggested \cite{2004MNRAS.348..187W,2005Natur.434..192F,2005ApJ...620L..43O,2006MNRAS.365..590K} but the possibility of some superluminous supernovae being genuine pair-instability supernovae (PISNe) from very massive stars beyond this limit \cite{2009Natur.462..624G,2012Sci...337..927G} and the suggestion of stars of up to $300\,\msun$ in the R136 star cluster in \tdor \cite{2010MNRAS.408..731C} have called this limit into question. Indeed, \cite{2014ApJ...780..117S} lifted some of the tension between the suggested $150\,\msun$ upper-mass limit and observational evidence for even more massive stars, and re-determined the upper-mass limit to be in the range of $200\text{--}500\,\msun$ from observations of the stellar content of the R136 star cluster. While the upper-mass limit is not the core question of this work, the extent of the IMF towards high masses affects estimates of massive star feedback in Sect.~\ref{sec:feedback}.

\begin{figure}
\centering
\includegraphics[width=0.7\textwidth]{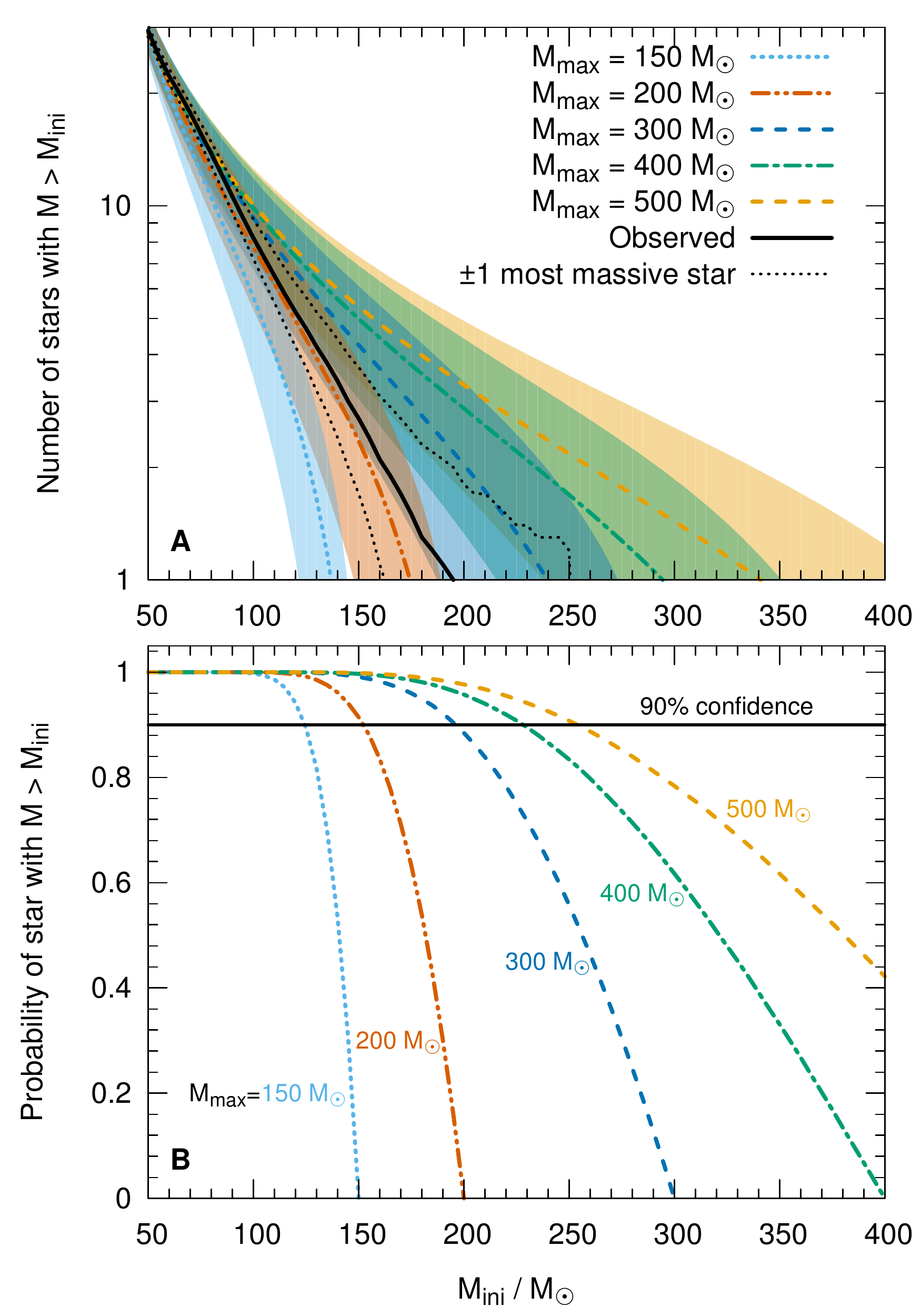}
\caption{\textbf{Number of stars (A) and probability that at least one star (B) is formed with a mass greater than $M_\mathrm{ini}$ for stellar upper-mass limits $M_\mathrm{max}$ of $150$, $200$, $300$, $400$ and $500\,\msun$.} In panel (A), also the observed number of stars are shown (black solid line) and the $1\sigma$ ranges of the expected number of stars due to stochastic sampling (shaded regions). Furthermore, we add and remove one star in the mass range $150\text{--}250\,\msun$ from the observed sample (black dotted lines in panel A) to indicate potential variability of the observed number of high-mass stars.}
\label{fig:upper-mass-limit}
\end{figure}

Observationally, it is difficult to obtain tight constraints on the upper-mass limit because of the paucity of very massive stars. Any attempt to do so therefore inevitability involves low-number statistics and must consider stochastic sampling. Here, we randomly sample stars from our inferred SFH and IMF until a sample of stars more massive than $15\,\msun$ is obtained that has the same total stellar mass as the observed 247 massive stars, \ie a sample that is compatible with our \tdor stellar sample in terms of sample size and selection criteria. We repeat the experiment 100,000 times and for different stellar upper-mass limits ($150$, $200$, $300$, $400$ and $500\,\msun$). From these Monte-Carlo experiments, we compute the number of stars and the probability of the formation of at least one star above certain initial masses (Fig.~\ref{fig:upper-mass-limit}). The given $1\sigma$ uncertainties of the number of stars above certain initial masses are the standard deviations of the 100,000 repetitions. These simulations implicitly assume that star formation is a stochastic process and that stars of any mass could have formed in \tdor. In reality this might not necessarily be true. Our sampling experiments do not account for the possibility of forming additional massive stars, \eg, by merging binary stars. This affects our ability to put strong constraints on the lower limit of the maximum birth mass of stars (see below).

Within $1\sigma$ uncertainties and a single-star framework, our \tdor stellar population is consistent with an upper-mass limit of $200\text{--}300\,\msun$ but not with an upper-mass limit of $150\,\msun$ (Fig.~\ref{fig:upper-mass-limit}A). VFTS\,1025 (also known as R136c), the most massive star in our sample, has an initial mass of $203^{+40}_{-44}\,\msun$ (Table~\ref{tab:stellar-parameters}). We therefore have no star above $\approx 250\,\msun$ in our sample. This constraint allows us to exclude an upper-mass limit of $\gtrsim 500\,\msun$ because we would otherwise expect to find at least one star initially more massive than $250\,\msun$ in $\gtrsim 90\%$ of the cases (Fig.~\ref{fig:upper-mass-limit}B). To further stress the importance of uncertainties due to stochastic sampling, we add and remove one star from the observations in the mass range $150\text{--}250\,\msun$; this mimics the hypothetical cases in which the very massive star Mk~42 was a genuine member of the VFTS and included in our sample or that the most massive star in our sample would have been found to be a binary and hence removed from our sample, respectively. In fact the apparently most massive star in our sample, VFTS\,1025, may be a wind-colliding binary given its strong X-ray emission \cite{2006AJ....131.2164T,2009MNRAS.397.2049S}. We only indicate the variability of the high mass end because (i) stochastic sampling is most important there and (ii) the high mass end puts the strongest constraints on the upper-mass limit.

The R136 star cluster has not been observed in the VFTS and is thus excluded from the discussion of the upper-mass limit here. However, it contains several very massive stars that would provide valuable information on the upper-mass limit. Table~8 in \cite{2016MNRAS.458..624C} provides a list of very massive stars ($\log L/\lsun \geq 6.2$) in \tdor and shows that our sample only includes one (VFTS\,1025) out of nine very massive stars in or around the R136 core region. As discussed in Sect.~\ref{sec:r136-imf}, we neither understand the stellar content of the R136 core yet nor the selection effects and observational biases from \cite{2016MNRAS.458..624C}, making it difficult to integrate the massive stars of this region into our discussion of the upper-mass limit. However, current evidence suggests that some of the stars in R136 may be initially as massive as $300\,\msun$ \cite{2010MNRAS.408..731C,2014A&A...565A..27H}, which would lead to yet larger values of a potential upper-mass limit. We also note that it is conceivable that the maximum birth mass of stars depends on the star formation conditions, \ie heating from previous stellar generations and/or low metallicities might increase the characteristic mass of stars and maybe also affect the maximum birth mass.

So far, we have sampled a population of single stars and are thus able to put conservative constraints on the largest possible upper-mass limit that can explain the stellar population of our \tdor sample. However, a sizeable fraction of massive stars will exchange mass with a binary companion during their lives \cite{2012Sci...337..444S,2013A&A...550A.107S}. Binary mass transfer can increase stellar masses (at most by a factor of 2) and thus needs to be accounted for when constraining the lowest possible upper-mass limit \cite{2014ApJ...780..117S} that can explain the \tdor massive star population. Unfortunately, because of the complex SFH in \tdor and challenging binary physics, a detailed population synthesis model is required to properly estimate a lower limit of the maximum birth mass of stars. We can therefore only suggest an effective stellar upper-mass limit of $200\text{--}300\,\msun$ that can explain the massive star population of \tdor within $1\sigma$ and a single-star framework. It is not possible to say whether a very massive star was born with its high mass or gained it later-on from binary mass transfer---a difference that does not matter when discussing, \eg, the present-day feedback of starburst stellar populations such as \tdor.

\section{Stellar feedback}\label{sec:feedback}

An IMF with a slope shallower than the Salpeter value that extends up to at least $200\,\msun$ as we find for \tdor has consequences for the feedback of stellar populations on their host galaxies. In this section we estimate the changes in stellar feedback by comparing the feedback of massive-star populations drawn from high-mass IMF slopes $\gamma<2.35$ with that from populations drawn from a Salpeter high-mass IMF slope ($\gamma=2.35$). We extend the high-mass IMFs down to $1\,\msun$ and follow an (Kroupa) IMF with a power-law exponent $\gamma=1.3$ below $1\,\msun$ down to $0.08\,\msun$ \cite{2001MNRAS.322..231K}. To facilitate comparisons of the feedback, we normalise to the total stellar mass of the population, \ie we consider the feedback per unit stellar mass. We emphasise that the following computations are estimates to understand the impact of a varying high-mass IMF slope and stellar upper-mass limit on stellar feedback that rely on simplifying assumptions (as described below). Our estimates give only an impression of the expected changes in stellar feedback for a varying high-mass end of the IMF.

Broadly speaking, stellar feedback can be divided into three categories: kinetic energy injected by stars via their winds and supernova explosions, ionising radiation and elemental abundance enrichment by the release of metals (\ie elements heavier than helium). The number of compact remnants left behind by massive stars is required to understand the rates of compact object mergers such as black hole binary mergers observed via their gravitational wave emission \cite{2016PhRvL.116f1102A,PhysRevLett.116.241103,PhysRevLett.118.221101}. For supernovae, we follow \cite{2003ApJ...591..288H} and make the following assumptions on the different explosion mechanisms and left over compact remnants depending on the initial mass $M_\mathrm{ini}$ of stars:
\begin{description}
\itemsep0.0em
\item[$9\leq M_\mathrm{ini}/\msun\leq 25$:] Stars explode as core-collapse supernovae (CCSNe), leaving neutron stars of mass $1.4\,\msun$ behind. 
\item[$25\leq M_\mathrm{ini}/\msun\leq 40$:] Stars explode as CCSNe but only give rise to a weak explosion and form $10\,\msun$ black-holes by fallback.
\item[$40\leq M_\mathrm{ini}/\msun\leq 100$:] Stars do not explode but directly collapse to black holes with masses equal to the stellar masses at the end of their lives.
\item[$100\leq M_\mathrm{ini}/\msun\leq 140$:] Stars explode in pulsational PISNe leaving a black-hole remnant of 30\% of the final stellar mass; 70\% of the final mass is assumed to be ejected in pulsationally-driven outbursts and the final core collapse supernova.
\item[$140\leq M_\mathrm{ini}/\msun\leq 260$:] Stars explode as PISNe, leaving no compact remnants behind.
\item[$M_\mathrm{ini}/\msun\geq 260$:] Stars become pair-unstable, but collapse to black holes rather than explode in a supernova.
\end{description}
The final stellar masses are taken from evolutionary models \cite{2011A&A...530A.115B,2015A&A...573A..71K}. Theoretically, PISNe are expected to only occur at low metallicities where stellar winds are weak enough to allow for the growth of large stellar cores \cite{2003ApJ...591..288H,2007A&A...475L..19L,2015A&A...573A..71K}. The very massive stars in \tdor are not expected to explode in PISNe \cite{2015A&A...573A..71K} although the exact details are a sensitive function of stellar wind mass losses. In our feedback estimates we nevertheless assume that stars with initial masses in the range $140\text{--}260\,\msun$ explode as PISNe to consider their impact in lower metallicity environments in the distant Universe. 

We estimate the wind feedback over a stellar life from the integrated wind momentum, $p_\mathrm{wind}=\dot{M} v_\infty \tau$, and wind energy, $E_\mathrm{wind}=0.5 \dot{M} v_\infty^2 \tau$, where $\dot{M}$ is the average wind mass-loss rate during a star's life, $v_\infty\propto v_\mathrm{esc}\propto \sqrt{M/R}$ the velocity of wind material at infinity which, for radiation-driven winds, is related to the escape velocity, $v_\mathrm{esc}$, from the surface of stars with mass $M$ and radius $R$, and $\tau$ the lifetime of stars. Except for the averaged wind mass loss, the stellar properties correspond to the zero-age main-sequence (ZAMS), non-rotating models of \cite{2011A&A...530A.115B,2015A&A...573A..71K}.

Regarding the release of metals, we consider metal production by CCSNe and PISNe at metallicities of $Z=0.001\text{--}0.002$. Because of the low metallicity, most metals produced in such stars are released through supernova explosions rather than stellar winds and we thus assume for simplicity that only those stars that explode contribute to the chemical abundance evolution of the host galaxy. For CCSNe, we use metal yields of $Z=0.002$ models \cite{1995ApJS..101..181W} and, for PISNe, metal yields of $Z=0.001$ models \cite{2014A&A...566A.146K}.

In terms of radiation feedback, we consider hydrogen (H~\textsc{i}) and helium (He~\textsc{ii}) ionising photons with wavelengths $\leq 91.2\,\mathrm{nm}$ and $\leq 22.8\,\mathrm{nm}$, respectively. The fraction of ionising radiation emitted by stars of given effective temperatures is estimated by assuming that stars behave like black bodies. The effective temperatures and bolometric luminosities are again taken from the ZAMS stellar models. The fraction of ionising radiation from black bodies is likely overestimating that of massive stars because \eg dense winds can re-absorb some of the ionising radiation and re-emit it at longer wavelengths. The effective temperatures of stars and hence the ionising radiation in our estimates depend---among other factors---on metallicity and age. Lower metallicities imply more ionising radiation because stars are hotter, whilst higher metallicities imply less ionising radiation. Older stars are cooler when they evolve towards the red supergiant branch significantly reducing the produced ionising radiation. We neglect ionising sources other than ZAMS stars, for example our estimates neither include the feedback from hot Wolf--Rayet stars nor X-ray binaries.

In Table~\ref{tab:feedback}, we summarise feedback enhancements from stellar populations drawn from an IMF with slope $\gamma=1.90$ compared to that of populations drawn from an IMF with a standard Salpeter slope of $\gamma=2.35$. At high-mass IMF slopes of $\gamma<2.00$, most of the stellar mass is found in massive stars and it is necessary to define an upper-mass limit to avoid a diverging total stellar mass. This implies that the upper mass cut influences feedback estimates and we therefore probe four different upper-mass limits, $M_\mathrm{max}$, of $150$, $200$, $300$ and $500\,\msun$. The limits are chosen to represent a realistic range of potential upper-mass limits as found in Sect.~\ref{sec:upper-mass-limit}.

The IMF in population synthesis calculations, galactic evolution models, large cosmological simulations etc. is often truncated at $100\,\msun$ \cite{2014MNRAS.444.1518V,2015MNRAS.446..521S,2015ApJ...805...20S}. To also illustrate the expected increase of stellar feedback in such situations, we consider the following two cases: (i) both the shallower and Salpeter IMFs extend up to a certain stellar upper-mass limit, $M_\mathrm{max}$, and (ii) only the shallower IMF extends up to $M_\mathrm{max}$ whereas the Salpeter IMF is truncated at $100\,\msun$. In the latter case, the increase in feedback is considerably more (Table~\ref{tab:feedback}).

\begin{figure}
\centering
\includegraphics[width=0.95\textwidth]{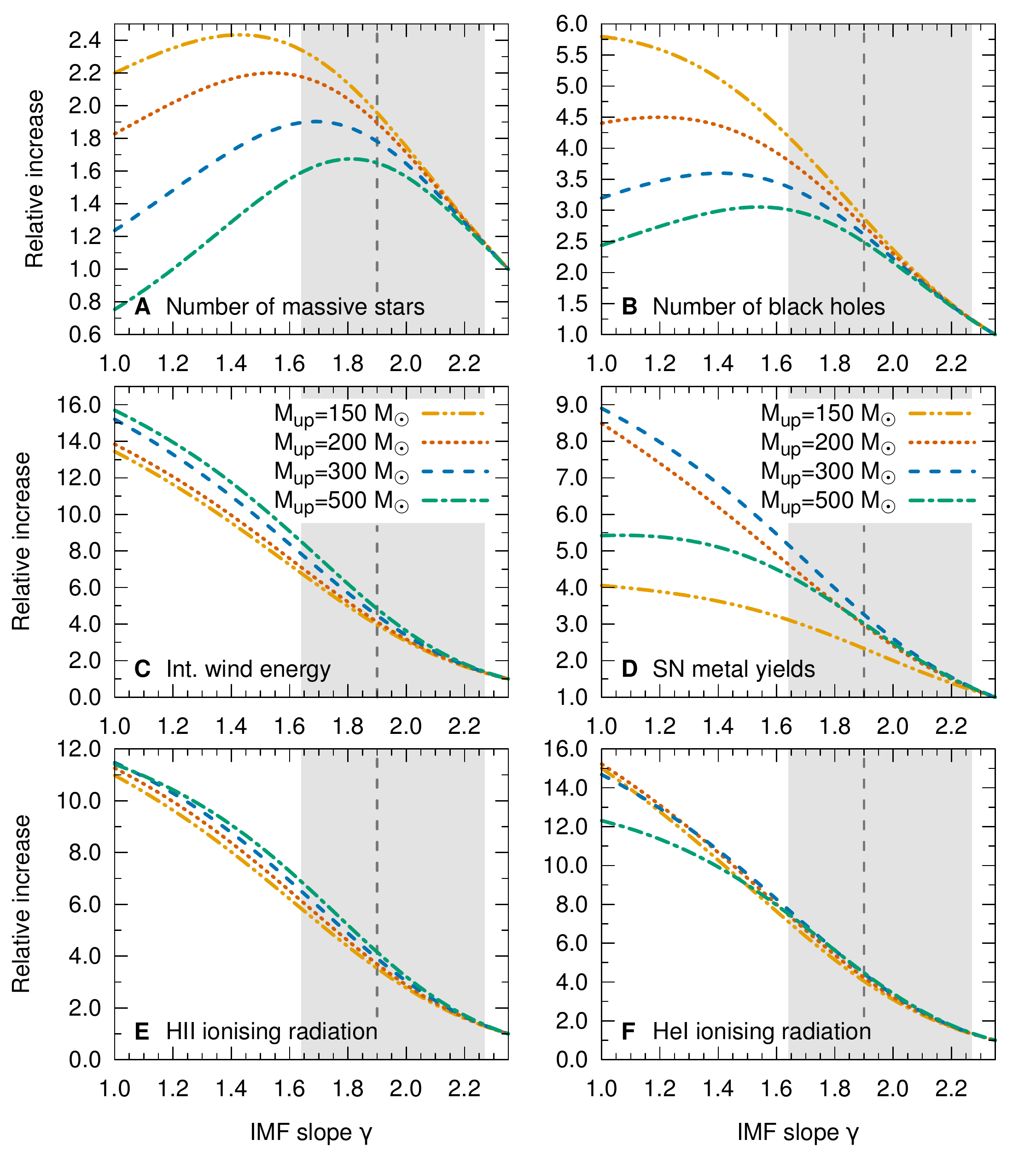}
\caption{\textbf{Relative increase of stellar feedback from massive stars with varying high-mass IMF slopes $\gamma$ and stellar upper-mass limits.} The reference points for the feedback are massive stellar populations born with a Salpeter IMF ($\gamma=2.35$). Shown are the increase in the (A) number of massive stars ($\geq 9\,\msun$), (B) number of black holes, (C) integrated wind energy, (D) supernova metal yields, (E) hydrogen (H~\textsc{i}) ionising radiation, and (F) helium (He~\textsc{ii}) ionising radiation. The provided values are estimates and, except for (A) and (B), rely on stellar models of certain metallicities and simplifying assumptions (see text for more details). The grey shaded regions indicate the IMF slope found for \tdor in this work.}
\label{fig:feedback}
\end{figure}

Later in Fig.~\ref{fig:feedback}, we also study changes in the estimated feedback because of variations in the IMF slope. These changes are larger than those from varying the upper mass limit when considering the $1\sigma$ range of our inferred IMF slopes. In the main text, we therefore report feedback variations because of different IMF slopes and a fixed upper mass limit of $200\,\msun$. In the following we discuss the changes stemming from different upper mass limits and IMF slopes separately.

\begin{table}
\caption{\label{tab:feedback}Stellar feedback enhancement. Ratios of the listed parameters for populations born with a high-mass IMF slope of $\gamma=1.90$ cf.\ $2.35$ (Salpeter). To facilitate comparison, the populations have the same total stellar mass and the IMFs are extended down to $0.08\,\msun$ using a Kroupa IMF \cite{2001MNRAS.322..231K}.}
\centering
\begin{tabular}{lcccccccc}
\toprule 
Upper-mass limit & \multicolumn{2}{c}{$150\,\msun$} & \multicolumn{2}{c}{$200\,\msun$} & \multicolumn{2}{c}{$300\,\msun$} & \multicolumn{2}{c}{$500\,\msun$}\\
Case & 1$^{*}$ & 1$^{\dagger}$ & 2$^{*}$ & 2$^{\dagger}$ & 3$^{*}$ & 3$^{\dagger}$ & 4$^{*}$ & 4$^{\dagger}$\\
\midrule
\midrule 
Stars with $M_\mathrm{ini}\geq 9\,\msun$ & 2.0 & 2.0 & 1.9 & 1.9 & 1.8 & 1.8 & 1.7 & 1.7\\
Stars with $M_\mathrm{ini}\geq 100\,\msun$ & 4.5 & -- & 4.5 & -- & 4.5 & -- & 4.4 & --\\
Core collapse SNe & 1.8 & 1.7 & 1.7 & 1.7 & 1.6 & 1.5 & 1.5 & 1.4\\
Black holes & 2.9 & 3.0 & 2.8 & 2.9 & 2.6 & 2.7 & 2.5 & 2.6\\
Black holes with masses $\geq 30\,\msun$ & 3.2 & 3.2 & 3.1 & 3.0 & 3.0 & 2.9 & 2.9 & 2.6\\
SN metal yields & 2.3 & 2.5 & 3.0 & 4.3 & 3.3 & 5.7 & 3.0 & 5.2\\
Integrated wind momentum & 4.7 & 9.1 & 4.7 & 12.3 & 4.8 & 19.4 & 5.0 & 29.9\\
Integrated wind energy & 4.0 & 6.9 & 4.1 & 9.6 & 4.5 & 16.0 & 4.8 & 25.3\\
ZAMS mass-to-light ratio & 0.30 & 0.23 & 0.29 & 0.18 & 0.27 & 0.14 & 0.25 & 0.11\\
Hydrogen (H\,I) ionising radiation & 3.5 & 5.0 & 3.7 & 6.5 & 3.9 & 8.6 & 4.2 & 11.3\\
Helium (He\,II) ionising radiation & 4.0 & 8.4 & 4.2 & 12.7 & 4.4 & 18.6 & 4.5 & 22.6\\
\bottomrule
\end{tabular}\\
\small
$^{*}$ Both the shallower and Salpeter IMFs extend up to $M_\mathrm{max}$\\ 
$^{\dagger}$ The shallower IMF extends up to $M_\mathrm{max}$ but the Salpeter IMF stops at $M_\mathrm{max}=100\,\msun$\\
\end{table} 

With an IMF slope of $\gamma=1.90$, the number of massive stars ($>9\,\msun$) increases by about 70\%--100\%, resulting in 50\%--80\% more core-collapse supernovae. The stellar feedback discussed above strongly depends on stellar mass with the most massive stars contributing strongest to the overall feedback (they have the strongest stellar winds and are the hottest and most luminous stars). This implies that stellar feedback is enhanced and that the mass-to-light ratio is decreased. In a ZAMS stellar population, the mass-to-light ratio is lowered by factors of about 3--4 depending on the upper-mass limit and up to factors of 4--10 when comparing to the case where the Salpeter IMF is truncated at $100\,\msun$. A lowered mass-to-light ratio directly affects inferred properties of unresolved stellar populations such as the star-formation rate.

The relative increase in the number of massive stars because of an IMF with a high-mass slope of $\gamma=1.90$ scales with mass according to ${\propto}M^{2.35-1.90}=M^{0.45}$, such that feedback from high-mass stars is more strongly enhanced than that from lower mass objects. For example, we find that the number of stars above $100\,\msun$ increases by a factor of about $4.5$, \ie more than the number of stars above $9\,\msun$. Such very massive stars may give rise to (pulsational) PISNe which would contribute greatly to the chemical abundance evolution of the Universe. Including PISNe, low-metallicity, massive stars are expected to roughly triple the metal production (without the PISNe contributions, the metal feedback roughly doubles). Compared to the case of an IMF truncated at $100\,\msun$, the metal production increases by up to a factor of 6.

The number of black holes increases by factors of 2.5--3.0, similar to the roughly tripled number of massive black holes ($\geq 30\,\msun$). These black-holes can be detected in binary black-hole mergers via their gravitational wave signal \cite{2016PhRvL.116f1102A,PhysRevLett.116.241103,PhysRevLett.118.221101}. The increase of black holes with an IMF shallower than the Salpeter IMF also translates into an increase of X-ray binaries which produce strong ionising radiation which we do not account for in our feedback estimates.

During their lives, massive stars have powerful winds and shape their surroundings. The integrated wind momentum and energy increase by factors of 4--5 and up to factors of 7--30 when compared to cases where the Salpeter IMF is truncated at $100\,\msun$. Massive stars are hot and their spectral energy distribution peaks in the ultra-violet such that they produce copious amounts of hydrogen (H~\textsc{i}) and helium (He~\textsc{ii}) ionising photons which early-on in the history of the Universe have contributed to its reionisation. For an IMF slope $\gamma=1.90$, we predict that the hydrogen ionising radiation from a population of LMC ZAMS stars increases by factors of 3.5--4.2 and up to factors of 5.0--11.3 if the Salpeter IMF is truncated at $100\,\msun$. The helium ionising radiation is even more strongly increased because it originates from hotter and hence more massive stars which are, relatively speaking, also more abundant (see above). The helium ionising radiation increases by factors of 4.0--4.5 and up to factors of 8.4--22.6 when compared to a truncated Salpeter IMF.

So far, we have only considered the case of an IMF slope of $\gamma=1.90$ for clarity. If the IMF slope indeed flattens because of heating from previous stellar generations in starbursts \cite{2005MNRAS.359..211L}, it is conceivable that the IMF slope may be even flatter in extreme starbursts such as those in the Antennae Galaxies or in the first generations of stars forming at low metallicity in the distant Universe \cite{2010AJ....140...75W,1999ApJ...527L...5B,2002Sci...295...93A}. To explore the whole range of potential IMF slopes given our inferred uncertainties, we calculate the change in the number of massive stars ($\geq 9\,\msun$) and black holes, and the increase in various stellar feedback in Fig.~\ref{fig:feedback} as a function of high-mass IMF slope, $\gamma$, and stellar upper-mass limit, $M_\mathrm{max}$.

As a function of IMF slope, the relative change in the number of massive stars ($\geq 9\,\msun$) and hence number of black holes reaches a maximum depending on the stellar upper-mass limit. The reason is that we study the increase in the number of stars and feedback per unit mass. For IMFs with $\gamma<2.00$, most of the total mass is in high-mass stars and not low-mass stars. The number of massive stars at fixed total population mass therefore drops for flatter IMFs and fixed upper-mass limit. Analogously, it also drops for fixed IMF slope and larger upper-mass limit. This drop is found at larger IMF slopes for more massive upper-mass limits. Figures~\ref{fig:feedback}A and~\ref{fig:feedback}B illustrate that the number of massive stars per unit mass and hence their feedback largely depend on the upper-mass limit for $\gamma \lesssim 2.00$ (as stated above). In particular, we expect that the rates of compact object mergers as seen via their gravitational wave emission might depend on the stellar upper-mass limit if stars are born with a top-heavy ($\gamma<2.00$) IMF. 

In contrast, the integrated wind momentum and ionising radiation are not that sensitive to the stellar upper-mass limit (Figs.~\ref{fig:feedback}C, \ref{fig:feedback}E and~\ref{fig:feedback}F). Allowing for larger maximum masses still results in more massive stars which produce individually more feedback but this is partly compensated by the decreasing number of massive stars for larger upper-mass limits. 

The increase in metal production from supernovae depends strongly on the upper-mass limit if PISNe contribute to the chemical enrichment. The strong dependence on the upper-mass limit in Fig.~\ref{fig:feedback}D is because PISNe only occur in stars with initial masses of about $140\text{--}260\,\msun$ and hence PISNe only contribute if the upper-mass limit allows for them. With an upper-mass limit of $M_\mathrm{max}=150\,\msun$, the PISN contribution is minimal and it is maximum for $M_\mathrm{max}=300\,\msun$. At even larger upper-mass limits, the number of massive stars decreases as does the metal production.

% table with all stellar parameters
\clearpage
\renewcommand{\arraystretch}{1.25}
\captionsetup{font=scriptsize}
%\begin{center}
\begin{landscape}
%\footnotesize
\scriptsize
\tabcolsep=0.15cm
\begin{ThreePartTable}
\renewcommand{\LTcapwidth}{\linewidth}
\begin{longtable}{llccccccccccc}
\caption{\label{tab:stellar-parameters}Stellar parameters for our sample stars. Tabulated are the observables (bolometric luminosity $\log L/\lsun$, effective temperature $T_\mathrm{eff}$, surface gravity $\log g$, projected rotational velocity $v\sin i$ and surface helium mass fraction $Y_\mathrm{obs}$) used to determine fundamental stellar parameters (initial mass $M_\mathrm{ini}$, initial rotational velocity $v_\mathrm{ini}$, age, present-day mass $M_\mathrm{present}$ and radius $R$) from stellar models \cite{2011A&A...530A.115B,2015A&A...573A..71K} using \bonnsai. The spectral types are from \cite{2011A&A...530A.108E,2014A&A...564A..40W,2015A&A...574A..13E}, and uncertainties are $1\sigma$ confidence levels if not stated otherwise. Luminosities marked with an (*) are predicted by the stellar models because they could not be inferred from observations. No fundamental stellar parameters are provided in cases where the models are unable to satisfactorily reproduce the observables (see text).}\\
\hline\hline
 VFTS & Spectral type & $\log L/\lsun$ & $T_\mathrm{eff}$ & $\log g$ & $v\sin i$ & $Y_\mathrm{obs}$ & Ref. & $M_\mathrm{ini}$ & $v_\mathrm{ini}$ & Age & $M_\mathrm{present}$ & $R$ \\
 No. & & & (K) & (cgs) & ($\mathrm{km}\,\mathrm{s}^{-1}$) & &  & ($\msun$) & ($\mathrm{km}\,\mathrm{s}^{-1}$) & (Myr) & ($\msun$) & ($\rsun$) \\
\hline
\endfirsthead
\caption{continued.}\\
\hline\hline
 VFTS & Spectral type & $\log L/\lsun$ & $T_\mathrm{eff}$ & $\log g$ & $v\sin i$ & $Y_\mathrm{obs}$ & Ref. & $M_\mathrm{ini}$ & $v_\mathrm{ini}$ & Age & $M_\mathrm{present}$ & $R$ \\
 No. & & & (K) & (cgs) & ($\mathrm{km}\,\mathrm{s}^{-1}$) & &  & ($\msun$) & ($\mathrm{km}\,\mathrm{s}^{-1}$) & (Myr) & ($\msun$) & ($\rsun$) \\
\hline
\endhead
\hline
\endfoot
001 & B1.5: V & $3.71^{+0.10}_{-0.10}$ & $20000^{+1000}_{-1000}$ & $3.80^{+0.10}_{-0.10}$ & $187^{+30}_{-30}$ & -- & \cite{B.stars} & $7.6^{+0.5}_{-0.5}$ & $210^{+79}_{-31}$ & $27.8^{+3.7}_{-3.5}$ & $7.6^{+0.5}_{-0.5}$ & $5.7^{+0.6}_{-0.6}$\\
003 & B1 Ia+ & $6.03^{+0.10}_{-0.10}$ & $21000^{+1000}_{-1000}$ & $2.50^{+0.20}_{-0.20}$ & ${<}58$ & -- & \cite{2015A&A...575A..70M} & $65.4^{+11.4}_{-10.9}$ & $70^{+104}_{-75}$ & $2.9^{+0.4}_{-0.3}$ & $56.4^{+10.3}_{-7.8}$ & $74.2^{+16.1}_{-9.2}$\\
004 & B2 V & $3.85^{+0.10}_{-0.10}$ & $20000^{+1020}_{-1020}$ & $4.00^{+0.15}_{-0.15}$ & $171^{+30}_{-30}$ & -- & \cite{B.stars} & $8.0^{+0.5}_{-0.4}$ & $200^{+90}_{-27}$ & $24.5^{+3.5}_{-3.3}$ & $8.0^{+0.5}_{-0.4}$ & $5.7^{+0.8}_{-0.6}$\\
005 & B2 V(n) & $3.82^{+0.10}_{-0.10}$ & $19000^{+1000}_{-1000}$ & $3.50^{+0.10}_{-0.10}$ & $198^{+30}_{-30}$ & -- & \cite{B.stars} & $8.0^{+0.4}_{-0.6}$ & $240^{+67}_{-40}$ & $29.2^{+3.9}_{-3.0}$ & $8.0^{+0.4}_{-0.6}$ & $7.7^{+0.9}_{-0.7}$\\
006 & Mid-late K & $3.46^{+0.20}_{-0.20}$ & $4100^{+150}_{-150}$ & -- & -- & -- & \cite{Cool.stars} & -- & -- & -- & -- & -- \\
007 & B1-2 V & $3.64^{+0.11}_{-0.11}$ & $20000^{+1210}_{-1210}$ & $3.70^{+0.12}_{-0.12}$ & $169^{+30}_{-30}$ & -- & \cite{B.stars} & $7.4^{+0.4}_{-0.6}$ & $200^{+83}_{-27}$ & $30.3^{+4.6}_{-4.2}$ & $7.4^{+0.4}_{-0.6}$ & $5.7^{+0.8}_{-0.6}$\\
008 & B0.5: V(n) & $4.17^{+0.14}_{-0.14}$ & $30000^{+2410}_{-2410}$ & $4.30^{+0.10}_{-0.10}$ & $241^{+30}_{-30}$ & -- & \cite{B.stars} & $13.2^{+1.3}_{-1.3}$ & $260^{+52}_{-44}$ & $0.1^{+2.9}_{-0.1}$ & $13.2^{+1.3}_{-1.3}$ & $4.5^{+0.4}_{-0.3}$\\
010 & B2 V & $3.50^{+0.10}_{-0.10}$ & $17000^{+1000}_{-1000}$ & $3.80^{+0.10}_{-0.10}$ & ${<}40.00$ & -- & \cite{B.stars} & $6.2^{+0.3}_{-0.4}$ & $40^{+9}_{-37}$ & $45.3^{+6.5}_{-5.0}$ & $6.2^{+0.3}_{-0.4}$ & $5.4^{+0.6}_{-0.5}$\\
\end{longtable}
\begin{tablenotes}
\small
\item (\emph{a}) 100\% CI, (\emph{b}) 73\% CI, (\emph{c}) 88\% CI, (\emph{d}) 70\% CI, (\emph{e}) 84\% CI, (\emph{f}) 72\% CI, (\emph{g}) 83\% CI, (\emph{h}) 87\% CI, (\emph{i}) 71\% CI, (\emph{j}) 75\% CI, (\emph{k}) 79\% CI, (\emph{l}) 82\% CI, (\emph{m}) 94\% CI, (\emph{n}) 81\% CI, (\emph{o}) 92\% CI
\end{tablenotes}
\end{ThreePartTable}
\end{landscape}
%\end{center}
% In total Ntot=572 stars, Nout=120 not reproduced by models

\end{document}